\documentclass[
 reprint,
 amsmath,amssymb,
 aps,
 pre
]{revtex4-2}

\usepackage{url}
\usepackage{amsmath,amssymb,graphicx,bbm}
\usepackage{mwe}
\usepackage{rotating}
\usepackage{subcaption}
\usepackage{calc}

\graphicspath{ {./figures/} }

\usepackage{graphicx}% Include figure files
\usepackage{dcolumn}% Align table columns on decimal point
\usepackage{bm}% bold math

\usepackage{paralist} % compact itemize
\DeclareMathOperator{\mc}{\enspace ,}

\DeclareMathOperator{\mf}{\enspace .}

\begin{document}

%\preprint{APS/123-QED}

\title{Estimating Peak-Hour Traffic Congestion Patterns For Interacting Agents On Urban Networks}

\author{Marco Cogoni}
\author{Giovanni Busonera}%
\author{Francesco Versaci}%
\affiliation{CRS4}%

\date{\today}

\begin{abstract}
  We study the emergence of congestion patterns in urban networks by modeling
  vehicular interaction by means of a simple traffic rule and by using a set
  of measures inspired by the standard Betweenness Centrality (BC). We
  consider a topologically heterogeneous group of cities and simulate
  the network loading during the morning peak-hour by increasing the
  number of circulating vehicles. At departure, vehicles are aware of
  the network state and choose paths with optimal traversal time.
  Each added path modifies the vehicular density and travel times for the
  following vehicles. Starting from an empty network and adding traffic
  until transportation collapses, provides a framework to study network's
  transition to congestion and how connectivity is progressively disrupted
  as the fraction of impossible paths becomes abruptly dominant.
  We use standard BC to probe into the instantaneous
  out-of-equilibrium network state for a range of traffic levels and
  show how this measure may be improved to build a better proxy for cumulative
  road usage during peak-hours. We define a novel dynamical measure to
  estimate cumulative road usage and the associated total time spent over the edges
  by the population of drivers. We also study how congestion starts with
  dysfunctional edges scattered over the network, then organizes itself into
  relatively small, but disruptive clusters.
\end{abstract}

\maketitle

\section{Introduction}
Urban networks have been widely studied in recent years~\cite{colak_understanding_2016,li_percolation_2015,kirkley_betweenness_2018}, both
their growth over time and their complex dynamics under different traffic
levels. Network science has considerably helped to improve our understanding
of cities and to analyze and predict the reaction of
the different parts of the network under stress~\cite{zeng_switch_2019,hamedmoghadam_percolation_2021}. Such predictive
analyses may be performed by considering just the geographical and
topological features of a city, without having to recur to experimental
traffic data, but these datasets constitute nonetheless the reference
against which every theoretical model should be evaluated~\cite{bongiorno_vector-based_2021,lee_morphology_2017}.
Urban networks belong to the special class of (almost) planar graphs and in
this context there have been some notable results
recently~\cite{diet_towards_2018,helbing_traffic_2001}.
The geographical embedding leads to several constraints
to topology such as limitations to the number of long-range connections
and to the maximum connectivity observed at the single edge level~\cite{aldous2013true}. 
The study of edge degree distributions did not improve much our understanding
of cities, but non-local higher-order metrics such as network centralities
have been widely used both for theoretical studies and for practical
applications with notable success~\cite{white_betweenness_1994,newman_measure_2005}.
One of the metrics that have been used the most in recent years is the
Betweenness Centrality (BC), which is defined as the total flow passing
over each node of the network when enumerating all Origin-Destination
(OD) pairs and connecting them via shortest-path routing.
This definition is easily extended to Edge BC (EBC) by counting edge usage instead.
We will write BC instead of EBC in the rest of paper since our focus will be on edges.
BC and other measures have been used to predict which edges are subject to
the highest traffic demand that leads to the breakup of the network into
functionally independent pieces. This phase transition has been well studied
via percolation theory and it is known that the size distribution of the resulting sub networks follows a power law with a critical exponent that depends
on traffic intensity and on the time of the day.
This transition has been observed for real world datasets in large cities
such as London, Beijing and New York City~\cite{cogoni2021stability,zeng_switch_2019,taillanter_empirical_2021}.
During the most extreme conditions, total network breakdown has been observed to last for several
hours or even days~\cite{hamedmoghadam_percolation_2021}.

The percolation phase transition is not specific of high congestion
levels, but an edge may be classified as dysfunctional even with vehicles
traveling just below the speed limit~\cite{li_percolation_2015,zeng_switch_2019,cogoni2021stability}: this transition simply signals a change in the network behavior that happens at all traffic levels, but for
different critical speeds. On the other hand, in this work we will follow the
more practical definition of calling an edge dysfunctional only when
it cannot receive traffic anymore, due to road density nearing the maximum
value and speed approaching zero~\cite{hamedmoghadam_percolation_2021}.
The transition happens when there is suddenly a large part of the desired
travel paths that are no longer usable.

Standard BC approaches can only grasp a limited picture of the network
under stress because BC relies on several assumptions~\cite{agryzkov2019variant,gao2013understanding}: traffic origins
and destinations uniformly spread over the nodes; shortest paths
with fixed cost function; non-interacting multiple paths sharing the same
edge; the amount of traffic density contributed by each vehicle is the same
regardless of its driving time.
Edge usage obtained from a BC computation approximates well a network
with very small (or extremely fast) agents, each using shortest path
navigation with no congestion awareness~\cite{kazerani2009can}. A urban
transportation network is characterized by vehicle travel times comparable
to the typical timescale of congestion buildup during peak-hours.

In this work we want to extend the standard BC as a measure of road congestion,
especially by taking into account the strong interaction effects observed in real traffic.
The interaction among vehicles depends on the duration and vehicular volume of the
network loading: the longer the edge travel times, the higher the probability of
finding a vehicle in a given road segment during the observation period. The present
approach stems directly from this fact and aims at focusing on the peak-hour periods,
that usually last for one hour, to estimate the cumulative traffic seen on the roads
during that finite time window. Thus, we propose a dynamical model to
compute the contribution to traffic at a road-segment level due to each vehicle
added to the network, while iteratively recomputing travel times after each addition.
From the vast literature on transportation we choose one of the simplest
models to describe vehicular behavior depending on geography (edge
properties) and on the dynamical network state: the single regime
Greenshields model~\cite{helbing_traffic_2001,rakha2002comparison},
for which speed starts at the free flow value to decrease linearly
to zero when maximal road density is attained.
To complement the traffic model, we assume that vehicles know exactly
the network state before their departure, in order to plan an optimal path.
The proposed scenario mimics a network with a mixture of self-driving cars
and human drivers generating a shortest-time route at the start of their trips.
It is likely that this will become increasingly relevant in the near future.

In this paper we employ our dynamical model and the associated metrics to
theoretically predict the behavior of five large cities under growing traffic.
Results will be validated against real traffic data in future works.
%%%%%%%%%%%%%%%%%%%%%%%%%%%%%%%%%%%%%%%%%%%%%%%%%%
\section{Methods}

\subsection{Interaction Model}
We aim at modeling the network evolution, as observed by travelers,
while the traffic increases from zero up to complete gridlock. The
desired network traffic over simulation time ($\tau$) will be added incrementally,
activating one new path $\pi(i)$ at each simulation step $i$.
Thus, $i$ can be interpreted both as the current number of
added paths and as a temporal marker to define the sequence of OD pairs
randomly generated for each simulation.

For simplicity, and to be able to compare results with the standard BC,
traffic will be added uniformly to the network. It is however straightforward
to adapt our procedure to any OD matrix.

We model the traffic network as a directed, weighted graph $G=(V,E)$
with $N=|V|$ nodes and $M=|E|$ edges. Each edge $e$ represents a
road segment between two intersections and it is characterized 
by three constant features: its physical length $l_e$, maximum speed
$v^*_e$ and number of lanes $c_e$.

The state of the network at each time step will be represented by 
the occupancy of vehicles added so far in each edge ${s_e(i)}$, where
$s_e(i) = \sum_{j=1}^i \sigma_e(j)$. We will define the occupancy $\sigma$ of a single
vehicle so that it will sum to one only when its total traveling time
is equal to or longer than the simulation time
(in general: $\sum_{e\in\pi(i)} \sigma_e(i) \le 1$).
To each edge we also associate a normalized vehicle density
$\rho_e(i)$, that increases monotonically as we add vehicles:
\begin{equation}
      \rho_e(i) = \frac{s_e(i) L}{l_e c_e} \in [0, 1] \mc
    \label{density_def}
\end{equation}
where $L$ is the average space occupied by one vehicle.

We define the occupancy $\sigma_e(i)$ of a new vehicle added to the
network using a simple approximation of the time it might spend on
edge $e$:
\begin{align}
  \sigma_e(i) &= \min\left(\frac{T_e(i)}{\tau}, 1\right) \mc
  & \text{where } T_e(i) &= \frac{l_e}{v_e(i)} \mc
                           \label{omega_def}
\end{align}
and with speed $v_e(i)$ following a Greenshields linear
law~\cite{jin2021introduction}:
\begin{equation}
  v_e(i) = v_e^*(1-\rho_e(i-1)) \mc
  \label{greenshields_def}
\end{equation}
thus the approximate time for the complete path $\pi(i)$ will be
$T_{\pi(i)}=\sum_{e\in\pi(i)}T_e(i)$.
A non-interacting system is obtained in the limit $L\rightarrow 0$.
In more detail, the state $s(i)$ of the network is obtained
iteratively ($s(i)=f(s(i-1))$), starting with an empty network
($s(0)=0$) and according to the following dynamic process:
\begin{itemize}
\item a pair of OD nodes is chosen, independently and uniformly at
  random, and the fastest path $\pi(i)$ connecting the nodes is computed;
\item starting from O, for each edge $e\in\pi(i)$, we accumulate the
  average occupancy $\sigma(i)$ induced by $\pi(i)$ during $\tau$:
  \begin{equation}
    s_e(i) = s_e(i-1) + \sigma_e(i) \mc
    \label{occup_update}
  \end{equation}
  and the number of vehicles on the edge:
  \begin{equation}
    n_e(i) = n_e(i-1) + 1 \mf
    \label{occup_update_int}
  \end{equation}
  Note that, as soon as the sum along $\pi(i)$ of the added $T_e(i)/\tau$ factors
  reaches $1$, we choose to skip the remaining edges up to D, to avoid
  adding more than a unit factor to vehicle occupancy along $\pi(i)$;
\item $v_e(i+1)$ and $T_e(i+1)$ are updated according to
  Eqs.~\ref{omega_def}~and~\ref{greenshields_def}, using the new
  $\rho_e(i)$ value;
\item this process is iterated until the desired total traffic is reached.
\end{itemize}

Intuitively, the occupancy factor induced by a vehicle over an edge
$e$ is proportional to the time the vehicle is supposed to spend
on it (${T_e}/{\tau}$), as forecast at the moment of its generation,
and the sum over the whole path will be equal to unity (certainty of
finding the vehicle within $\pi$ during $\tau$) only when
$T_\pi\ge\tau$.
The initial OD pairs find a nearly empty network, so their fastest paths
and travel times are very similar to the $L=0$ case. With rising
traffic, however, edges slow down and subsequent fastest paths will be
slower and, in general, different. Some edges will eventually reach
maximum density and become dysfunctional. If the fastest route from
origin to destination comprises a dysfunctional edge (i.e., the network
is disconnected) we still choose to add the initial part of the path, but
we will skip the edges starting from the first dysfunctional one.
This allows us to model the backward propagation of traffic jams
observed at high traffic
volumes~\cite{olmos_macroscopic_2018,taillanter_empirical_2021}.
Since the order in which OD pairs are added to the network can lead to
different final results, we replicate the system and perform several
simulations to analyze the stability of the results. 

The main limitations of this model are: the Greenshields model is very rough;
no explicit time evolution as in the Nagel-Schreckenberg CA model, but just
a sequence of path addition in a \emph{first come, better served} approach;
no true time evolution means that long paths added initially will generally
experience faster times (uncongested) for edges that will likely become
congested later by newer paths; OD pairs are uniformly extracted at random
over the network: it is known, e.g., that morning and evening peak hours show
an opposite average traffic direction~\cite{colak_understanding_2016}.

\subsection{Cumulative BC definition}
BC implicitly assumes that the different paths do not
interact, i.e., that the cost (typically the shortest length between
two points) does not change when adding agents in the network. If,
however, we assume that the active network traffic does alter the cost
function, as it is the case when choosing fastest vs shortest paths,
then the standard BC gives a biased picture of bottlenecks and
hotspots. A good approximation of the BC can be
obtained by sampling the OD pairs uniformly~\cite{riondato2016fast},
and it is natural to think about it as a dynamical process of
subsequent path additions. In particular, if the exact paths, their order
of appearance and their travel times are known, the timescale $\tau$
of the dynamical phenomenon becomes a crucial parameter to better 
estimate the edge visiting frequencies.

Several studies extending the original BC concept have been presented
in the past~\cite{bavaud_interpolating_2012,newman_measure_2005,white_betweenness_1994},
to deviate from the idea of shortest paths by introducing routing randomness.
Here, on the other hand, we extend the BC idea by taking into account the evolution of
the network state. To this aim, we define a Cumulative BC (CBC), based on the average
occupancy $s_e(i)$, defined in Eq.~\ref{occup_update} as:
\begin{equation}
  \label{eq:CBC}
  \gamma_e(i) = \frac{s_e(i)}{i}\mf
\end{equation}
This quantity is normalized and tends to the original BC for $L\rightarrow0$.

\section{Simulation details}
We select the vehicular transportation layer of the urban networks relative to five large cities and their surroundings from OpenStreetMap. The radius $R$ of the circle inscribed in each region varies from $12$ km for Rome to $20$ km for Boston, and the number of edges of the corresponding graphs ranges from about 65k
of Rome to about 200k of London. For every city, we perform $10$ simulations, each with a reshuffled order of the OD additions to the network to estimate the sensitivity of the evolving configurations to external conditions. The total number of added ODs was $10^6$, sufficient to bring all cities to a deeply congested state.
The total computation time for simulation and analysis is less than one day
for each city, running with shared-memory parallelism on Intel E5-2680 nodes
(12 cores, 2 threads/core) with 250~GB RAM.

\section{Results and discussions}
\subsubsection{Transition to fragmented network}
\begin{figure}[h!]
\label{fig_rejected}
    \begin{subfigure}{\linewidth}
    \centering
      \begin{subfigure}{\linewidth/5 - 0.3em}
        \includegraphics[width=\linewidth]{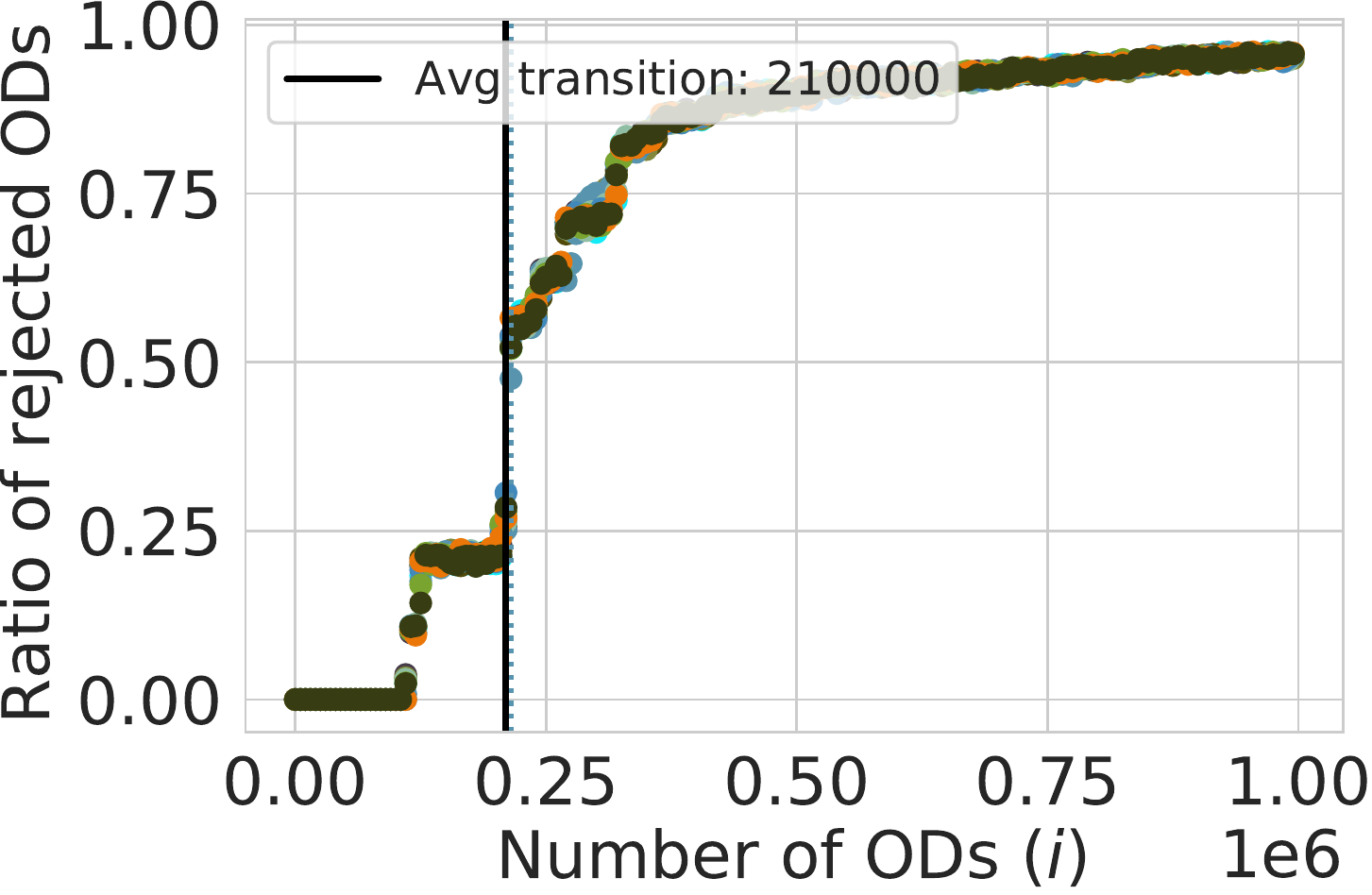}
        \caption{Nairobi}
        \label{fig:rejected_Nairobi}
      \end{subfigure}
      \begin{subfigure}{\linewidth/5 - 0.3em}
        \includegraphics[width=\linewidth]{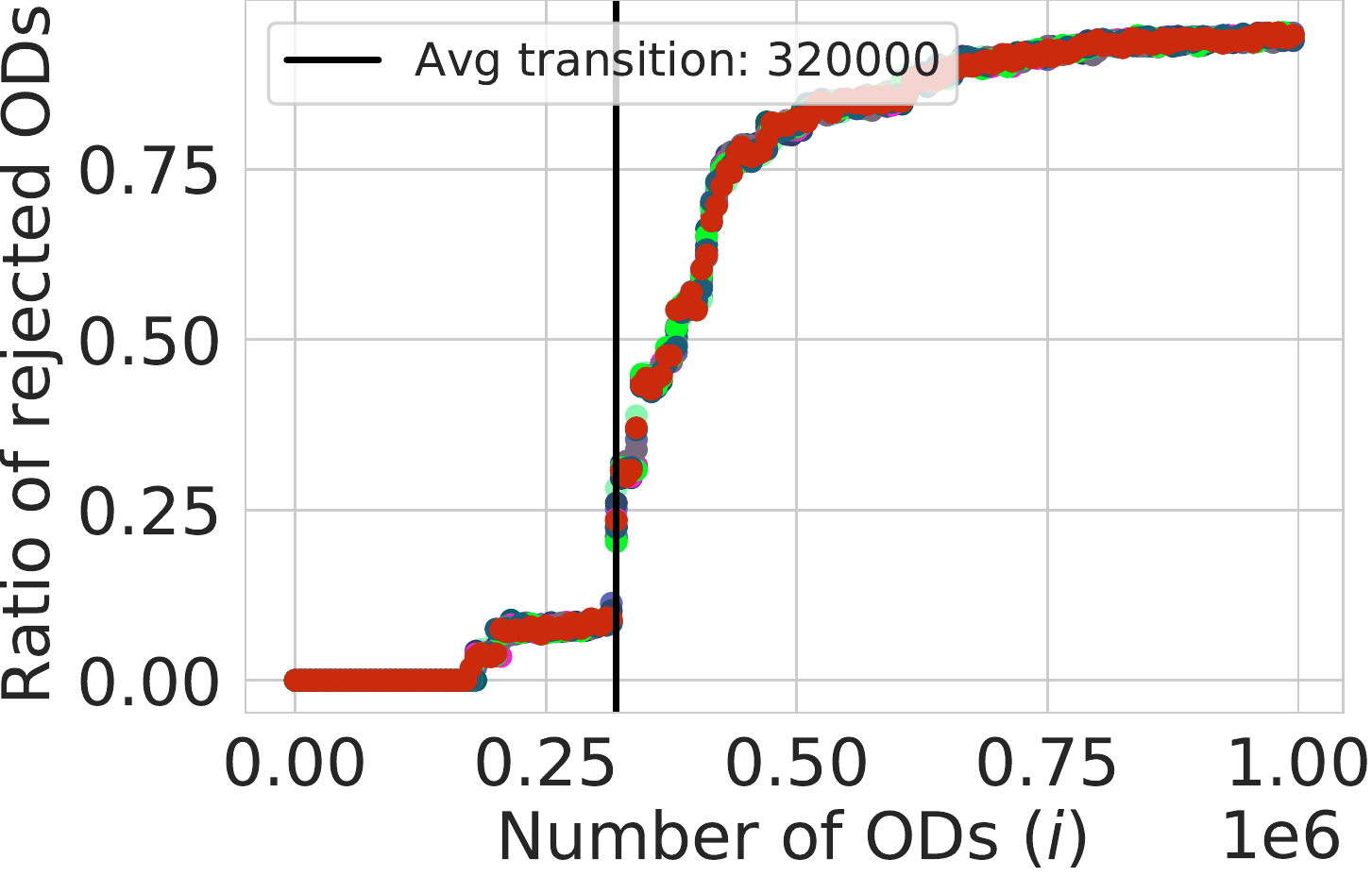}
        \caption{Rio}
        \label{fig:rejected_Rio}
      \end{subfigure}
      \begin{subfigure}{\linewidth/5 - 0.3em}
        \includegraphics[width=\linewidth]{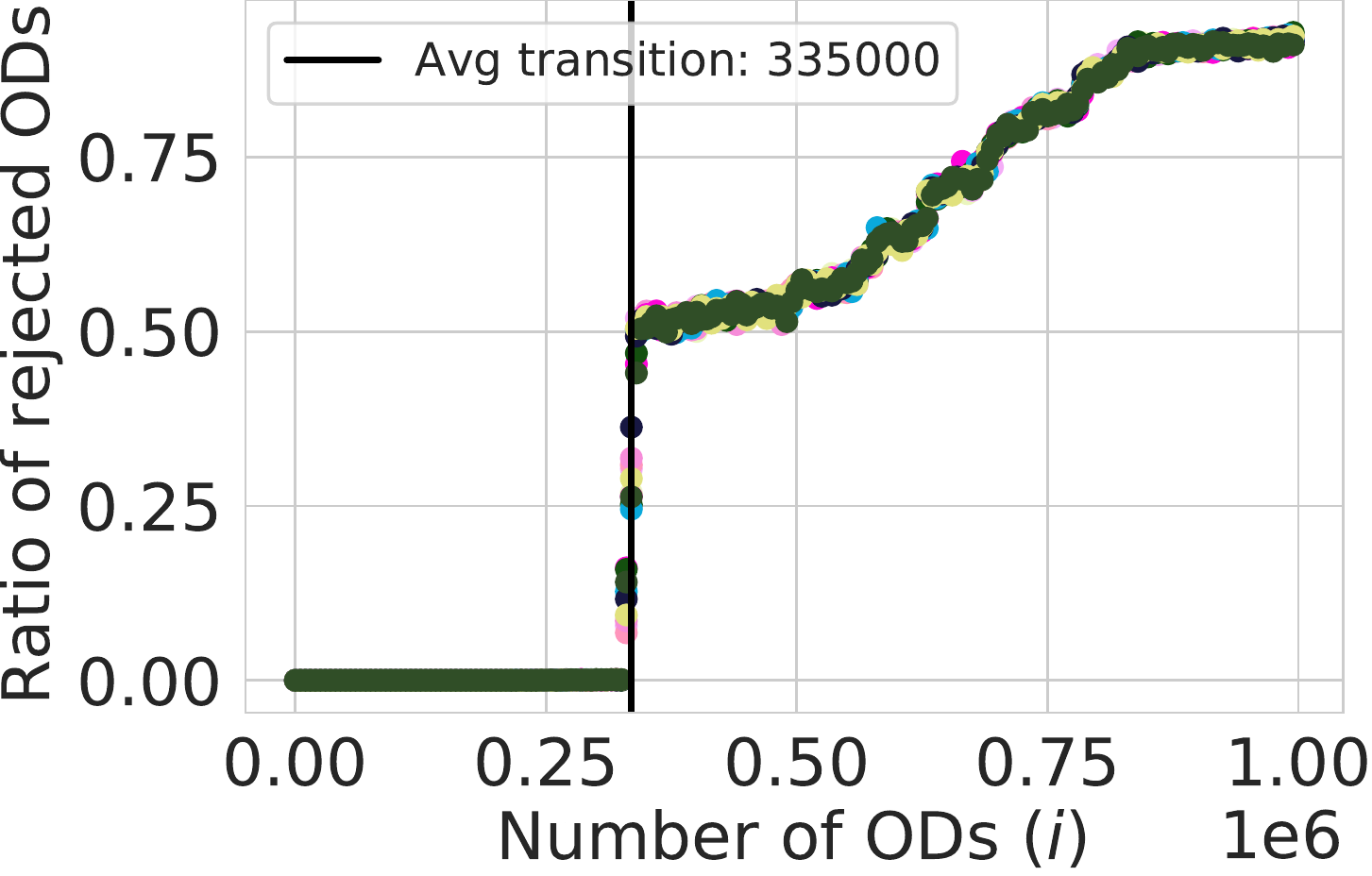}
        \caption{London}
        \label{fig:rejected_London}
      \end{subfigure}
      \begin{subfigure}{\linewidth/5 - 0.3em}
        \includegraphics[width=\linewidth]{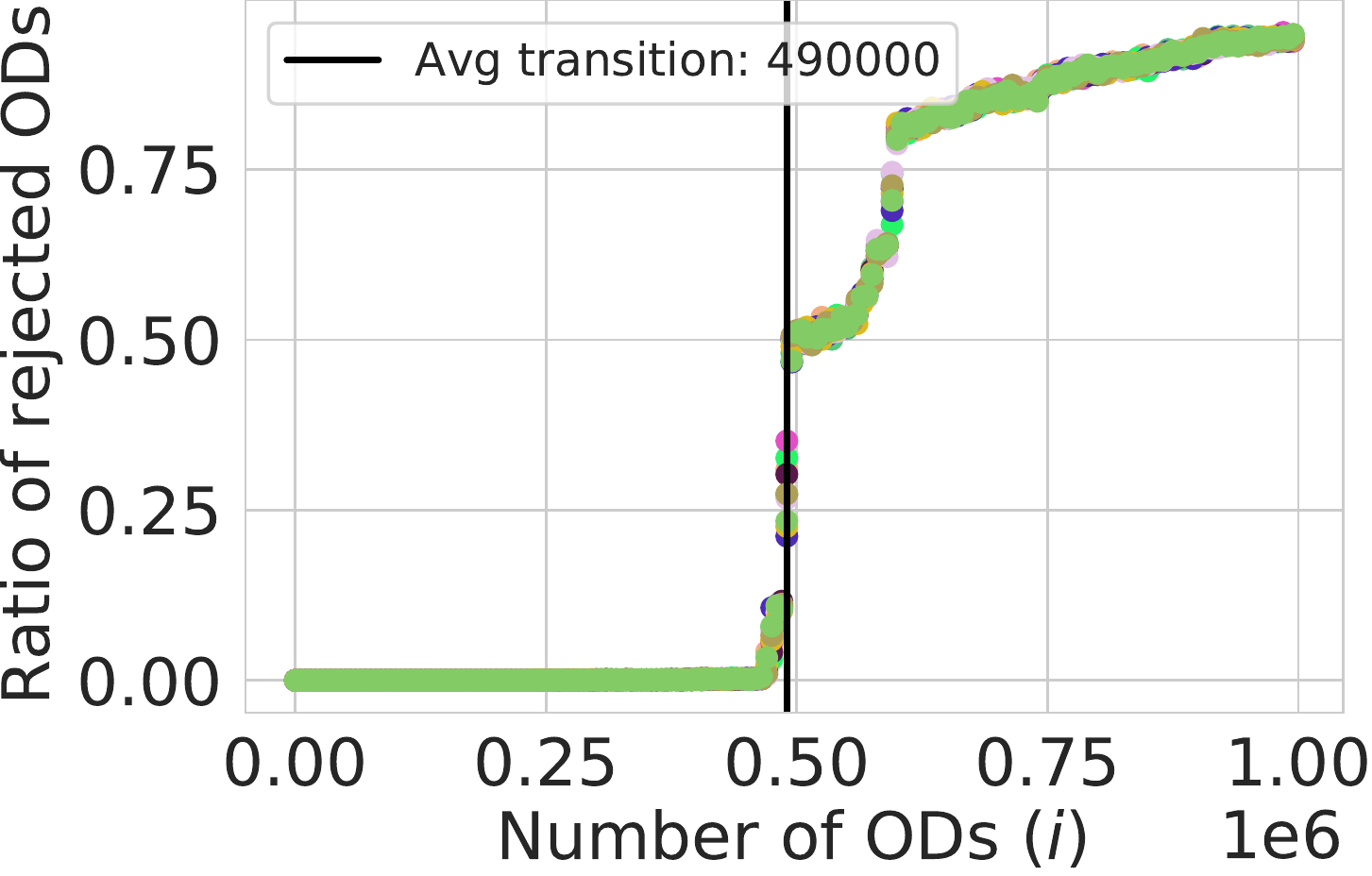}
        \caption{Rome}
        \label{fig:rejected_Rome}
      \end{subfigure}
      \begin{subfigure}{\linewidth/5 - 0.3em}
        \includegraphics[width=\linewidth]{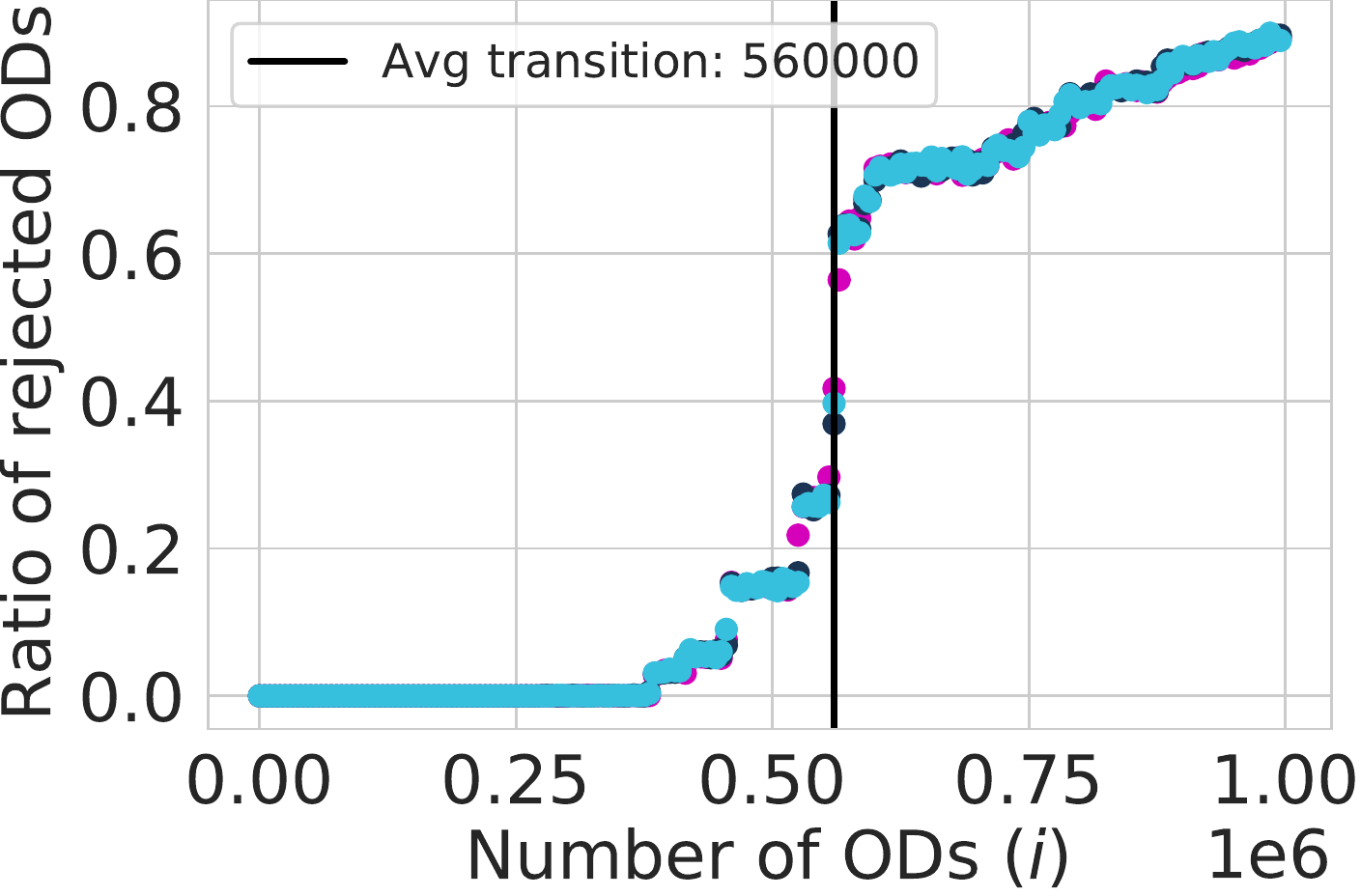}
        \caption{Boston}
        \label{fig:rejected_Boston}
      \end{subfigure}
    \end{subfigure}
\caption{Fraction of (partially) rejected paths for increasing traffic ($i$)}
\label{fig:rejected_ods}
\end{figure}

\begin{figure}[h!]
    \begin{subfigure}{\linewidth}
    \centering
      \begin{subfigure}{\linewidth/2 - 0.5em}
        \includegraphics[width=\linewidth]{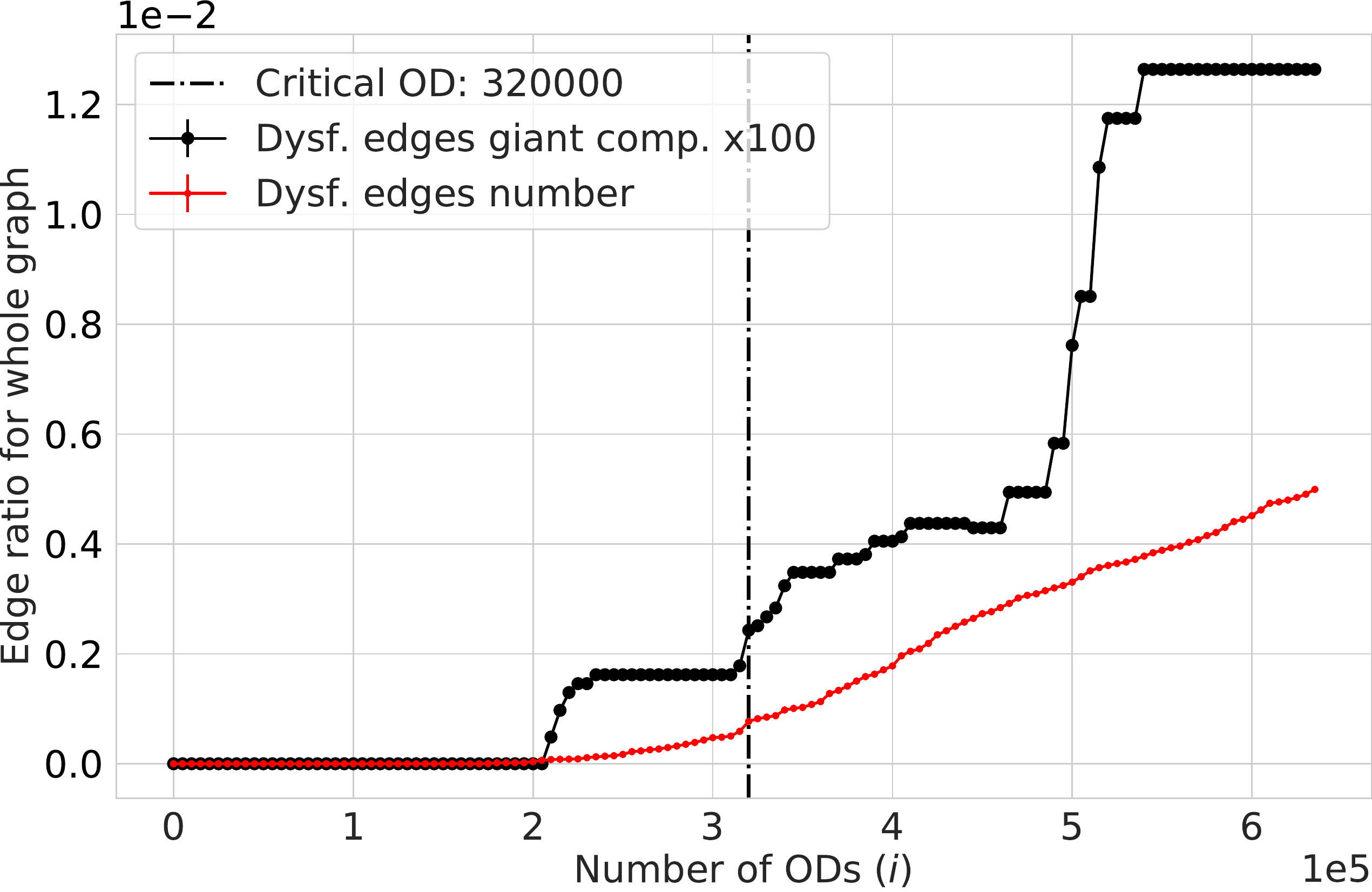}
      \end{subfigure}
      \begin{subfigure}{\linewidth/2 - 0.5em}
        \centering
        \begin{tabular}[t]{l@{\hskip .4in}c@{\hskip 0.2in}c}
        \hline
        & \# dysf. & LC size\\ 
        \hline
        \noalign{\vskip 1mm}
        Boston & $2\cdot10^{-3}$ & $4\cdot10^{-5}$  \\
        London & $5\cdot10^{-4}$ & $2\cdot10^{-5}$  \\
        Rome & $3\cdot10^{-3}$ & $6\cdot10^{-5}$    \\
        Rio & $1\cdot10^{-3}$ & $3\cdot10^{-5}$     \\  
        Nairobi & $1\cdot10^{-3}$ & $9\cdot10^{-5}$ \\
        \hline
        \end{tabular}
      \end{subfigure}
    \end{subfigure}
\caption{Spatial correlation of congested edges. Left: Fraction of dysfunctional edges (w.r.t.\ the entire graph) found at the transition in Rio as traffic increases in red, and in black (magnified $100$ times), the associated largest connected component cluster size fraction. Right: Values of red ($\#$ dysf.) and black (LC size) curves at the transition for each city.}
\label{fig:dysf_edges}
\end{figure}

\begin{figure}[h!]
\label{fig_cumul_length_time}
    \begin{subfigure}{\linewidth}
    \centering
      \begin{subfigure}{\linewidth/2 - 0.5em}
        \includegraphics[width=\linewidth]{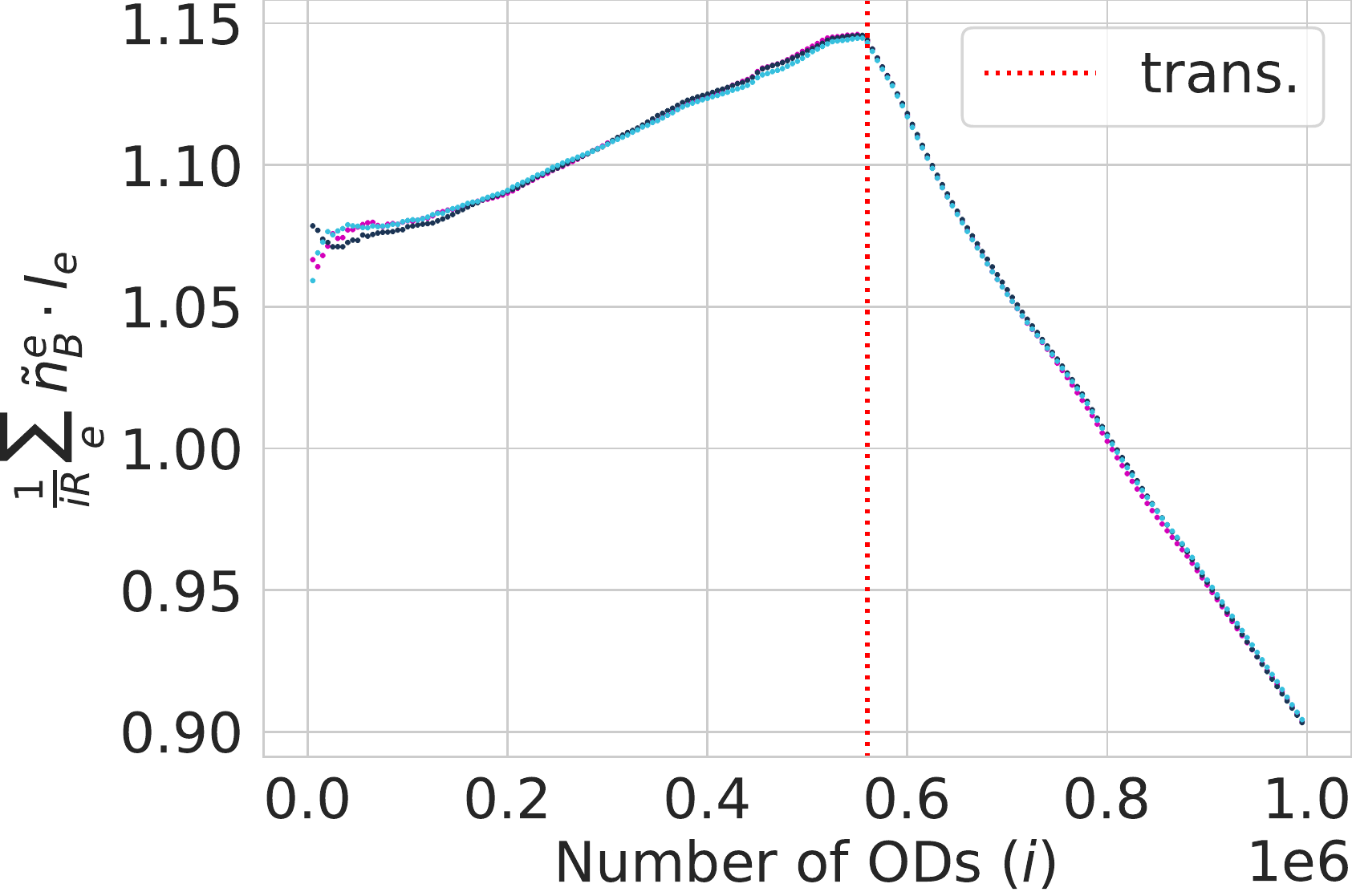}
        \label{fig:cumul_length}
      \end{subfigure}
      \begin{subfigure}{\linewidth/2 - 0.5em}
        \includegraphics[width=\linewidth]{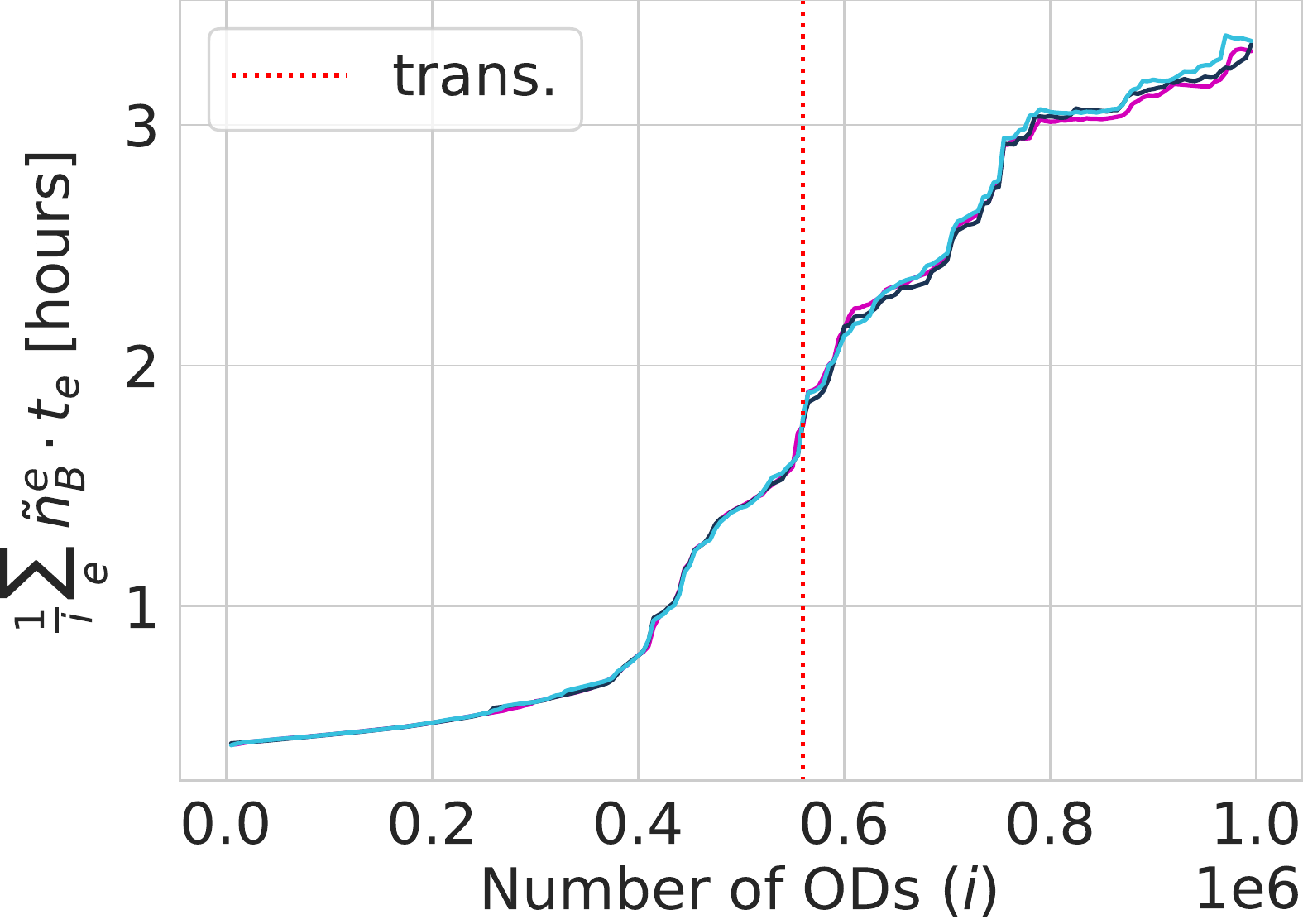}
        \label{fig:cumul_time}
      \end{subfigure}
    \end{subfigure}
\caption{(Average path length ratio $\mathbf{D}(i)$ (with respect to $R$) (left) and path travel time $\mathbf{T}(i)$ (right) per vehicle for Boston. Each color represents a different replica with reshuffled OD order.}
\label{fig:cumuls}
\end{figure}

We first focus on the behavior of the network while it receives an increasing number
of paths. Since edges become progressively dysfunctional as $\rho\rightarrow1$, at some
traffic level (set by $i$) the graph becomes disconnected, which
means that some OD pairs will belong to functionally separated subgraphs. When this occurs,
all paths connecting those OD pairs will include a dysfunctional edge and we say that the OD pair is 
(at least partially) rejected. This is a coarse indicator of a transition in the network behavior.
We present rejection ratio curves in Fig.~\ref{fig:rejected_ods}: Nairobi (Fig.~\ref{fig:rejected_ods}a) and Rio (Fig.~\ref{fig:rejected_ods}b) are the first to
collapse at $i\approx 210$k and $i\approx 320$k, respectively, with moderately steep curves.
The different transition types depend upon the topological features of each city (see maps on Fig.~\ref{fig_matrix1}): London (Fig.~\ref{fig:rejected_ods}c) and Rome (Fig.~\ref{fig:rejected_ods}d), being traversed by rivers, abruptly break at $i\approx 335$k and $i\approx 490$k, respectively, leaving both cities split in two halves ($50\%$ of rejected ODs). The collapsed bridges of both cities, due to heavy usage, are clearly visible in Fig.~\ref{fig_matrix1}(c). Once broken in two parts, London is able to better support local traffic better than Rome that collapses quickly even for shorter ODs, as visible from the different slopes after the transition. Boston (Fig.~\ref{fig:rejected_ods}e) is apparently the most resilient to breakup, with the first hints of congestion appearing at $i\approx 560$k, and a global behavior similar to Nairobi and Rio.
\onecolumngrid\
\twocolumngrid\
\newpage
We now shift our focus to the spatial configuration of dysfunctional edges: to display their typical growth over the graph, we choose Rio in Fig.~\ref{fig:dysf_edges} (left): the number of dysfunctional edges (red curve) and that of edges belonging to the Largest strongly connected dysfunctional Cluster (LC) grow almost monotonically with $i$ (black curve). It is worth noting that, although scattered dysfunctional edges appear early and grow monotonically, the LC lags and shows a staircase-like behavior, indicating that dysfunctional edges take time to coalesce into a set of clusters. We measure only the LC size and the plateaus are explained by new dysfunctional edges appearing in distant parts of the city. Remarkably, the table on the right of Fig.~\ref{fig:dysf_edges} shows that, for all cities, a small number of dysfunctional edges of order $0.1\%$ of the total edges, is enough to severely hinder transportation efficiency.
\onecolumngrid\
\twocolumngrid\
In order to further analyze the transition to congestion, we define two additional measures:
the average path length ratio: $\mathbf{D}(i) = \frac{1}{iR} \sum_e n_e(i)\cdot l_e$ and the
average path time per vehicle: $\mathbf{T}(i) = \frac{1}{i} \sum_e n_e(i)\cdot t_e$.
Since $n_e(i)$ is the number of vehicles that chose edge $e$ in $\tau$, by summing its products with edge lengths (traversal times) we obtain the global distance run (time spent) by all vehicles in $\tau$. $\mathbf{D}(i)$, with respect to the traffic level $i$, shows different regimes up to the transition: a constant or gentle slope for London and Rio, respectively, a bit steeper for Boston; a marked peak at the transition for all cities except Nairobi and Boston (Fig.~\ref{fig:cumuls}); a final decrease, for all cities after the transition, mainly due to partial truncation of paths added after the transition. $\mathbf{T}(i)$ per vehicle are similar for all cities: Fig.~\ref{fig:cumuls} (right) shows a gentle slope for low traffic regimes, which suddenly steepens nearing the transition with the dysfunctional edges restricting the choice of optimal paths.
For brevity, $\mathbf{D}(i)$ and $\mathbf{T}(i)$ are shown in Fig.~\ref{fig:cumuls} only for Boston: as traffic rises, avoidance of congested roads leads to an increase in the average travel length, but in general by less than $\approx10\%$.
\onecolumngrid\
\twocolumngrid\
\subsubsection{BC at increasing levels of traffic}
Standard BC may be used to probe the network state, after the addition of a number of vehicles, to obtain instantaneous information on how congestion affects path choices for increasing traffic levels. Fig.~\ref{fig_matrix1} (a) and (b) show how the BC maps of our five cities change from the empty state to the transition. Specifically, in Fig.~\ref{fig_matrix1} (b), edges with either larger or lower BC with respect to the empty state, are shown in red and blue respectively. It is worth to note that the roads with the highest BC with little traffic become dysfunctional long before the transition as entirely different path choices emerge: blue roads in Fig.~\ref{fig_matrix1} (b) are those that lost most of their flow due to early saturation while the red ones became important at higher congestion levels.

\subsubsection{Cumulative BC}
The ``loaded'' standard BC discussed above, does not describe the network behavior during the whole simulation time, but just what would happen to a small batch of vehicles added on top of a system with a specific traffic load. The Cumulative BC (CBC), on the other hand, takes into account the total contributions of all vehicles added within $\tau$. The CBC maps are shown in Fig.~\ref{fig_matrix1}(c) exactly at the transition $i_c$ specific for each city,
so each edge carries the contribution to the global traffic experienced during the whole simulation time. Several edges show high CBC values not detected as important by a BC in Fig.~\ref{fig_matrix1} (a). This is especially true for peripheral roads that become important only when the fastest options are already saturated: notably, Boston shows a somewhat busy eastern region
and Rome has several radial roads that show a relatively large usage.
    
\subsubsection{Total Time Spent in Traffic (TS)}
\begin{figure}[h!]
    \begin{subfigure}{\linewidth}
    \centering
      \makebox[0pt]{\rotatebox[origin=c]{90}{
        (a) Standard BC
      }\hspace*{2em}}%
      \begin{subfigure}{\linewidth/5 - 0.5em}
        \includegraphics[width=\linewidth]{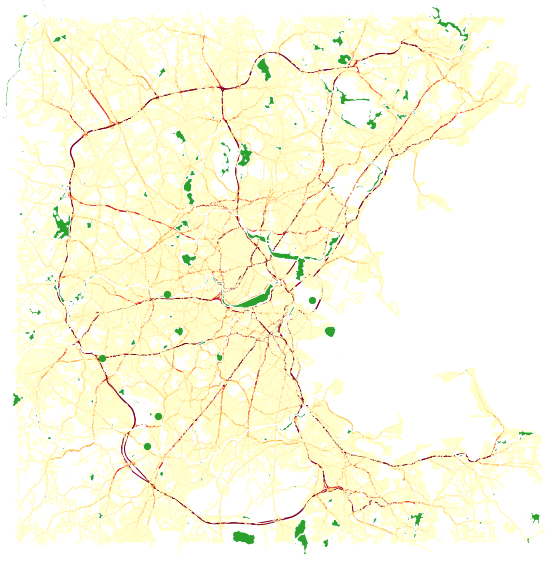}
      \end{subfigure}
      \begin{subfigure}{\linewidth/5 - 0.5em}
        \includegraphics[width=\linewidth]{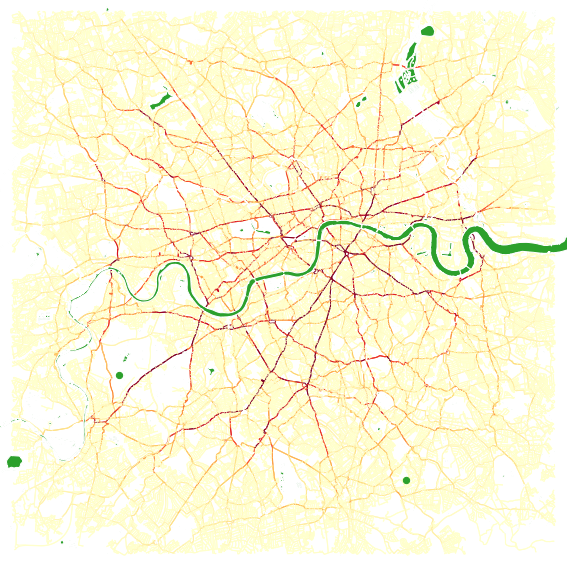}
      \end{subfigure}
      \begin{subfigure}{\linewidth/5 - 0.5em}
        \includegraphics[width=\linewidth]{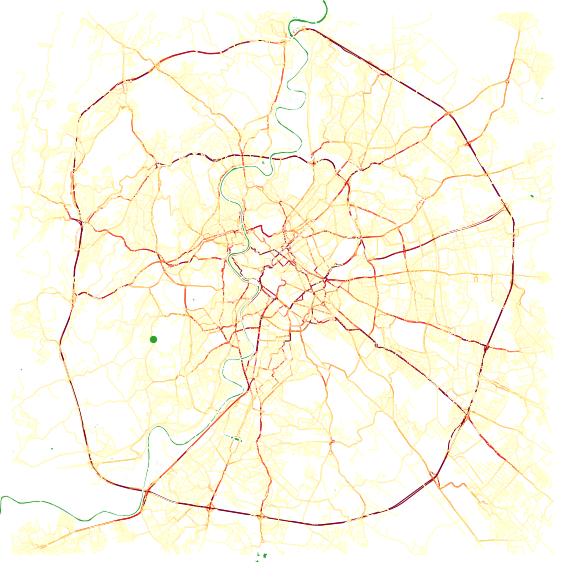}
      \end{subfigure}
      \begin{subfigure}{\linewidth/5 - 0.5em}
        \includegraphics[width=\linewidth]{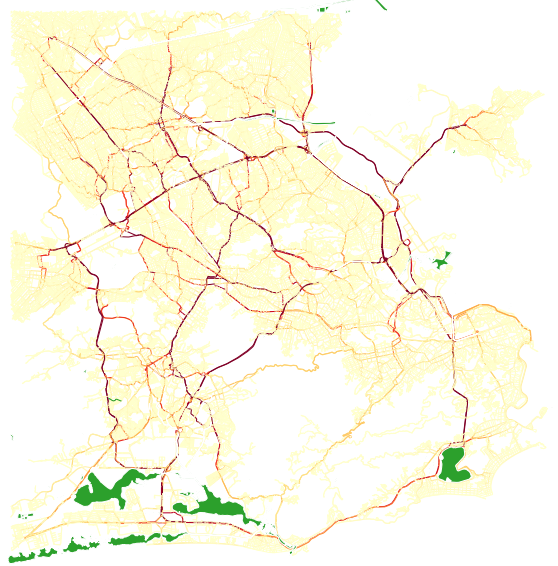}
      \end{subfigure}
      \begin{subfigure}{\linewidth/5 - 0.5em}
        \includegraphics[width=\linewidth]{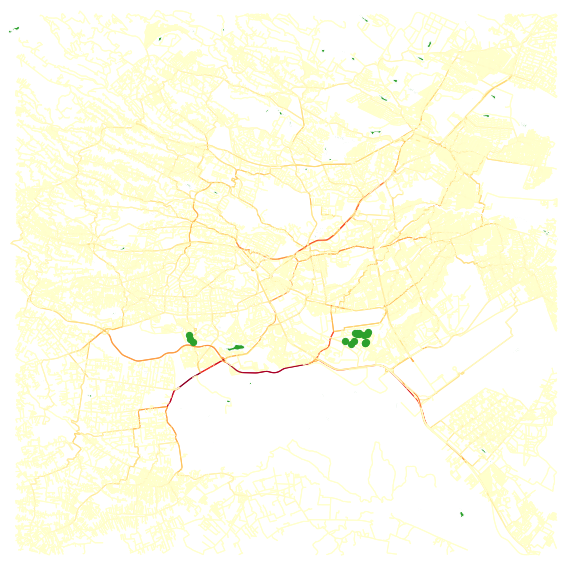}
      \end{subfigure}
    \end{subfigure}
    
    \par
    \begin{subfigure}{\linewidth}
    \centering
      \makebox[0pt]{\rotatebox[origin=c]{90}{
        (b) $\Delta BC$
      }\hspace*{2em}}%
      \begin{subfigure}{\linewidth/5 - 0.5em}
        \includegraphics[decodearray={-1.5 1.0 -1.5 1.0 -1.5 1.0},width=\linewidth]{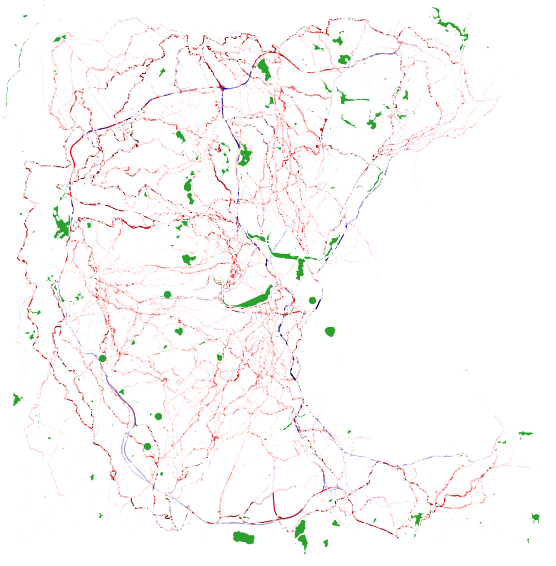}
      \end{subfigure}
      \begin{subfigure}{\linewidth/5 - 0.5em}
        \includegraphics[decodearray={-1.5 1.0 -1.5 1.0 -1.5 1.0},width=\linewidth]{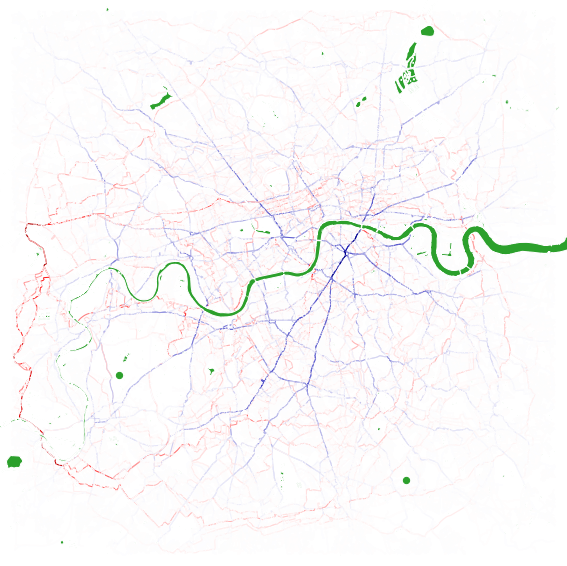}
      \end{subfigure}
      \begin{subfigure}{\linewidth/5 - 0.5em}
        \includegraphics[decodearray={-1.5 1.0 -1.5 1.0 -1.5 1.0},width=\linewidth]{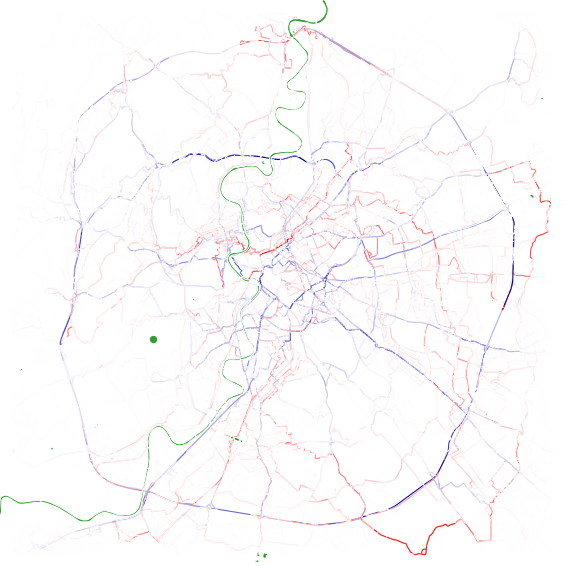}
      \end{subfigure}
      \begin{subfigure}{\linewidth/5 - 0.5em}
        \includegraphics[decodearray={-1.5 1.0 -1.5 1.0 -1.5 1.0},width=\linewidth]{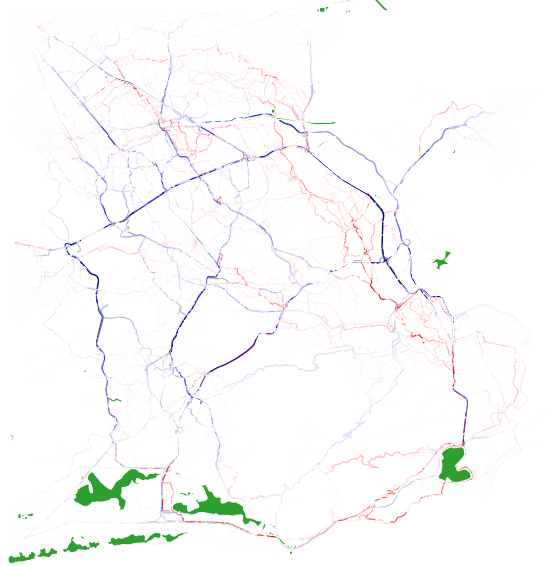}
      \end{subfigure}
      \begin{subfigure}{\linewidth/5 - 0.5em}
        \includegraphics[decodearray={-1.5 1.0 -1.5 1.0 -1.5 1.0},width=\linewidth]{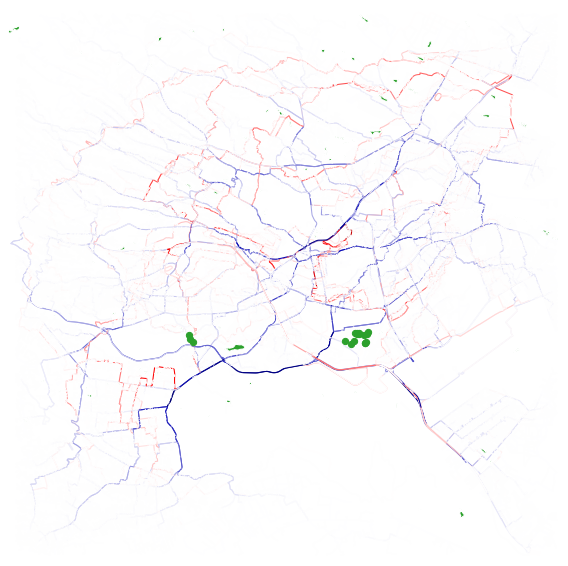}
      \end{subfigure}
    \end{subfigure}
    
    \par
    \begin{subfigure}{\linewidth}
    \centering
      \makebox[0pt]{\rotatebox[origin=c]{90}{
        (c) CBC
      }\hspace*{2em}}%
      \begin{subfigure}{\linewidth/5 - 0.5em}
        \includegraphics[width=\linewidth]{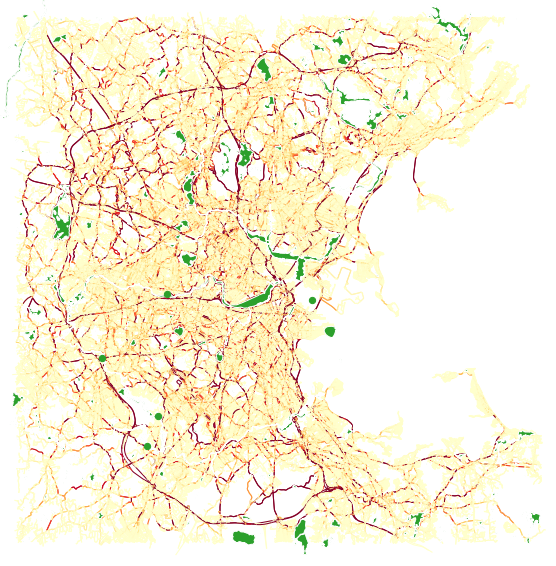}
      \end{subfigure}
      \begin{subfigure}{\linewidth/5 - 0.5em}
        \includegraphics[width=\linewidth]{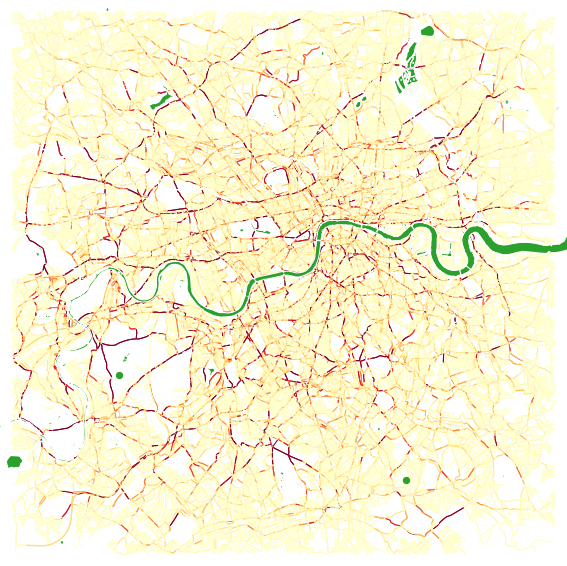}
      \end{subfigure}
      \begin{subfigure}{\linewidth/5 - 0.5em}
        \includegraphics[width=\linewidth]{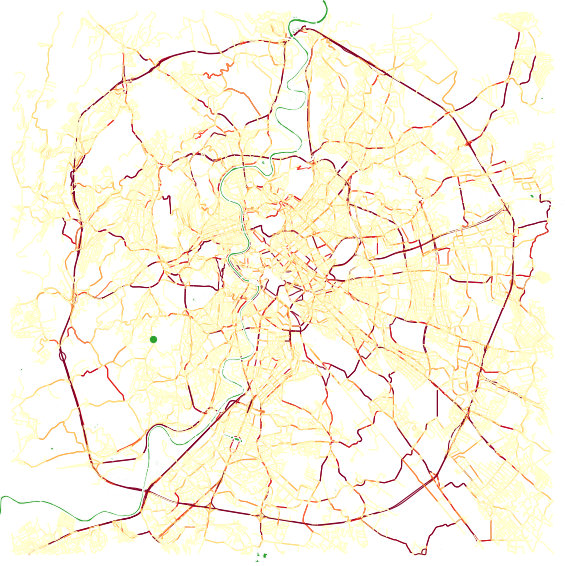}
      \end{subfigure}
      \begin{subfigure}{\linewidth/5 - 0.5em}
        \includegraphics[width=\linewidth]{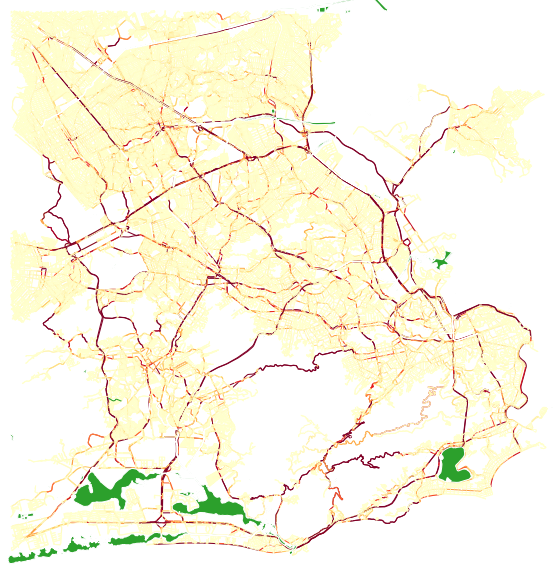}
      \end{subfigure}
      \begin{subfigure}{\linewidth/5 - 0.5em}
        \includegraphics[width=\linewidth]{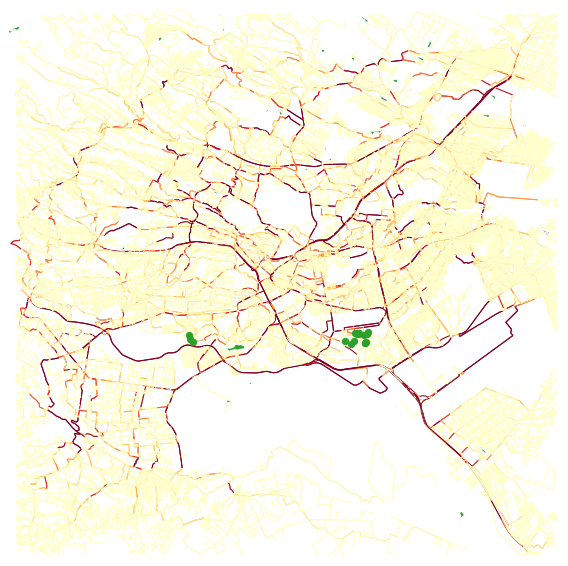}
      \end{subfigure}
    \end{subfigure}
    
        \par
    \begin{subfigure}{\linewidth}
    \centering
      \makebox[0pt]{\rotatebox[origin=c]{90}{
        (d) Time Spent
      }\hspace*{2em}}%
      \begin{subfigure}{\linewidth/5 - 0.5em}
        \includegraphics[width=\linewidth]{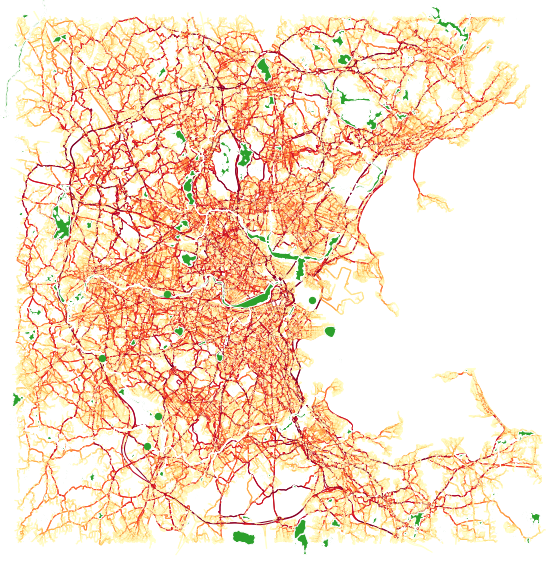}
      \end{subfigure}
      \begin{subfigure}{\linewidth/5 - 0.5em}
        \includegraphics[width=\linewidth]{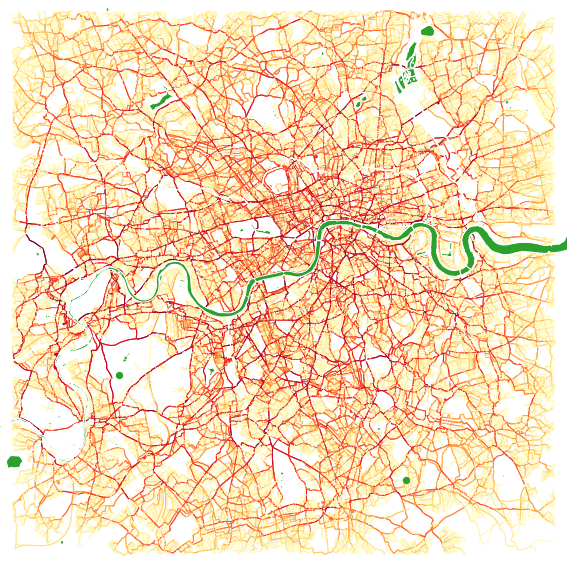}
      \end{subfigure}
      \begin{subfigure}{\linewidth/5 - 0.5em}
        \includegraphics[width=\linewidth]{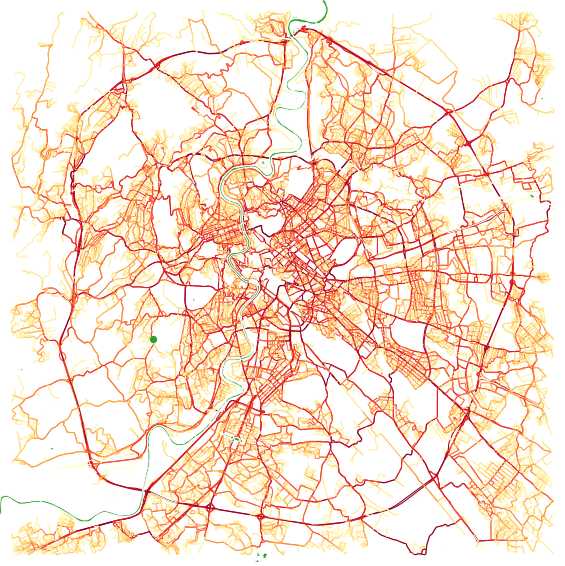}
      \end{subfigure}
     \begin{subfigure}{\linewidth/5 - 0.5em}
        \includegraphics[width=\linewidth]{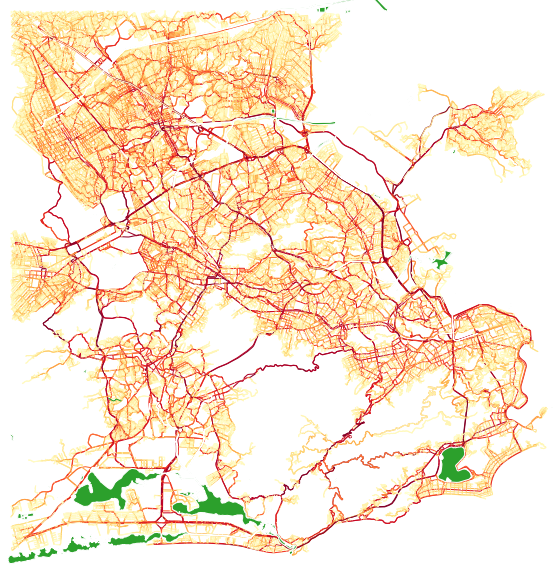}
      \end{subfigure}
      \begin{subfigure}{\linewidth/5 - 0.5em}
        \includegraphics[width=\linewidth]{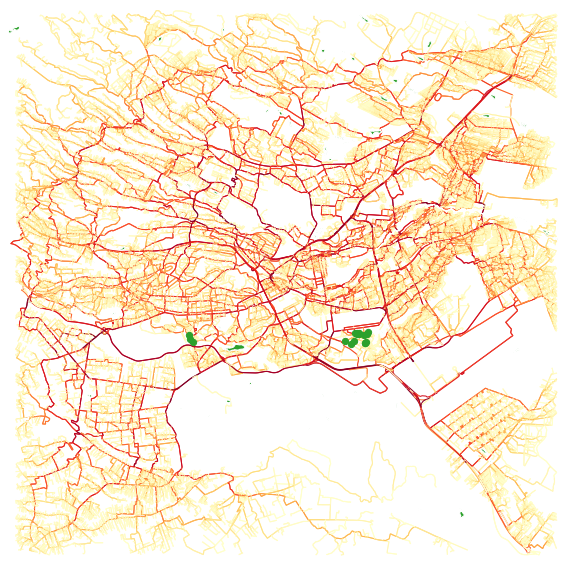}
      \end{subfigure}
    \end{subfigure}
    
    \par
    \begin{subfigure}{\linewidth}
    \centering
      \setcounter{subfigure}{0}%
      \renewcommand\thesubfigure{\roman{subfigure}}
      \begin{subfigure}{\linewidth/5 - 0.5em}
        \caption{Boston}
      \end{subfigure}
      \begin{subfigure}{\linewidth/5 - 0.5em}
        \caption{London}
      \end{subfigure}
      \begin{subfigure}{\linewidth/5 - 0.5em}
        \caption{Rome}
      \end{subfigure}
      \begin{subfigure}{\linewidth/5 - 0.5em}
        \caption{Rio}
      \end{subfigure}
      \begin{subfigure}{\linewidth/5 - 0.5em}
        \caption{Nairobi}
      \end{subfigure}
    \end{subfigure}
\caption{\small(a): Standard BC maps for all cities: red edges are within the $99$-th percentile of the BC distribution for the empty network. (b): $\Delta$BC at the transition: deep red and dark blue edges increase or decrease their BC by $\approx 100\%$, respectively. (c): red edges show the $99$-th percentile of the CBC distribution at the transition. (d): red edges represent the $99$-th percentile of the Time Spent in Traffic distribution at transition (per unit length, in order not to visually overestimate longer edges).}
\label{fig_matrix1}

\end{figure}

After defining the CBC as a proxy for the number of vehicles to be observed during the network loading, we can use this information to obtain some interesting features, such as the total time spent by all vehicles on each edge:
$\mathbf{TS}_e(i) = \frac{1}{l_e} \sum_e n_e(i) T_e(i)$.
This measure provides an intuitive way of estimating the total wait experienced, per unit length, during $\tau$ by all vehicles sharing an edge. As visible in Fig.~\ref{fig_matrix1}(d), the maps show the relative importance of each road: low values belong to seldom used roads while high values emerge for roads that are frequently chosen whether with an empty network (coinciding with BC) or at near-congestion. In general, urban highways, with several lanes and higher speed limits, are
the primary candidates to display top scores in this metric, but as congestion grows, other roads, often not planned for heavy use, start to attract a large fraction of vehicles and may compete with highways. It should be noted how most of the top-ranked roads in London are bridges over the Thames that in fact are the first to become dysfunctional (as seen in Fig.~\ref{fig:rejected_London}), leading to a city split in two roughly equal parts.

% figure matrix for BC, deltaBC, top99 CBC and TS
\onecolumngrid\
\twocolumngrid\

\subsubsection{CBC and total time spent in traffic}
\begin{figure}[h!]
    \begin{subfigure}{\linewidth}
    \centering
      \makebox[0pt]{\rotatebox[origin=c]{90}{
        (a) CBC
      }\hspace*{2em}}%
      \begin{subfigure}{\linewidth/5 - 0.5em}
        \includegraphics[width=\linewidth]{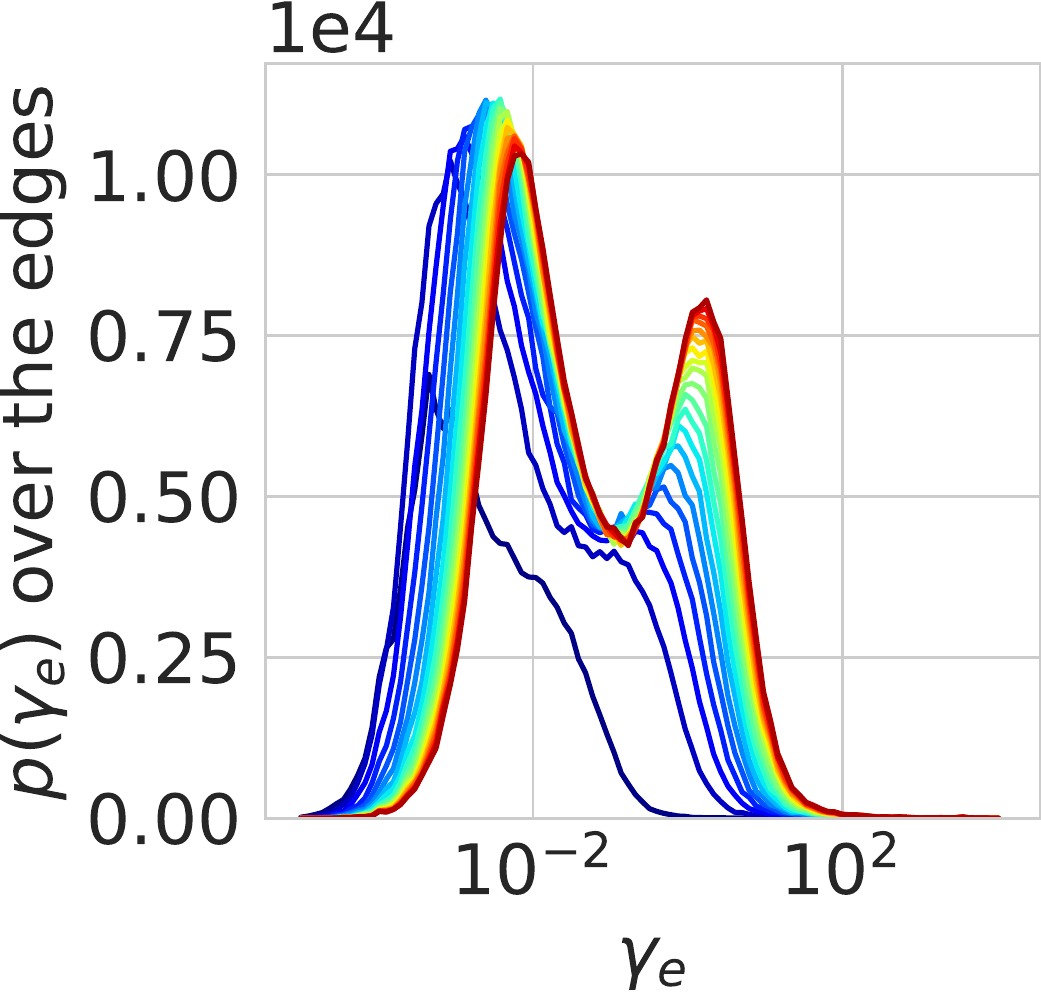}
      \end{subfigure}
      \begin{subfigure}{\linewidth/5 - 0.5em}
        \includegraphics[width=\linewidth]{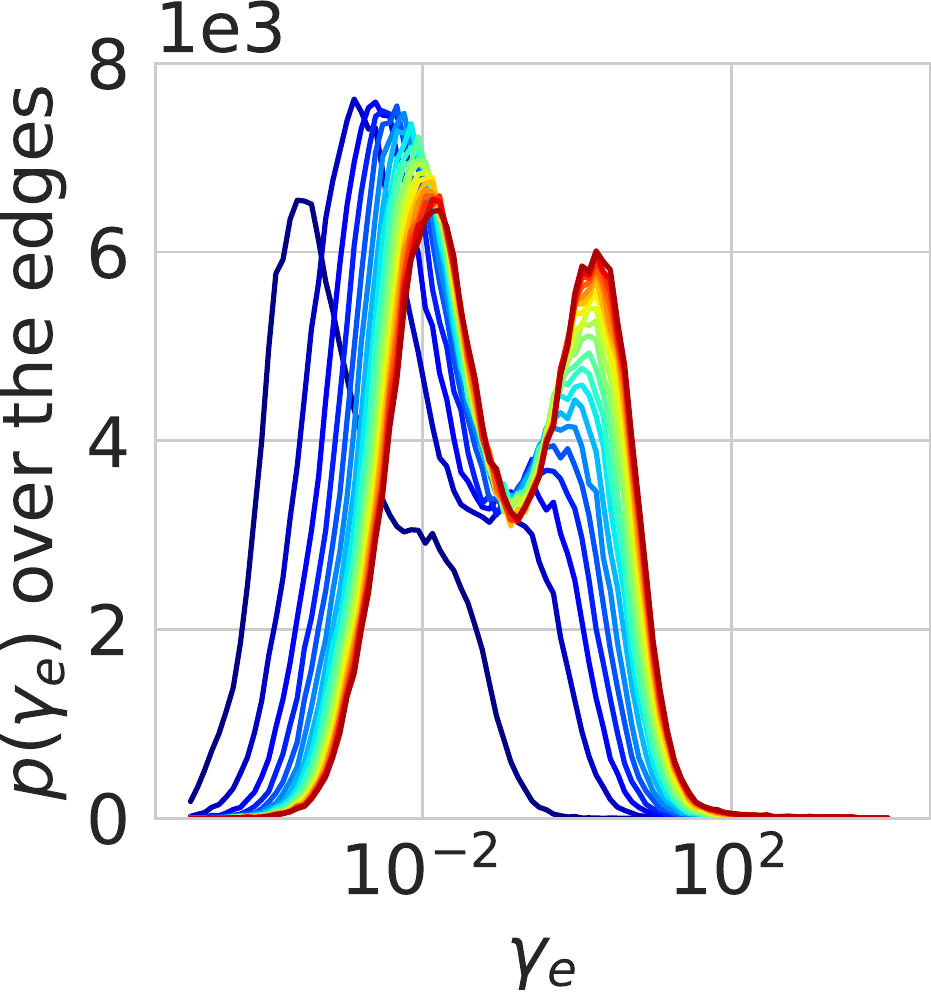}
      \end{subfigure}
      \begin{subfigure}{\linewidth/5 - 0.5em}
        \includegraphics[width=\linewidth]{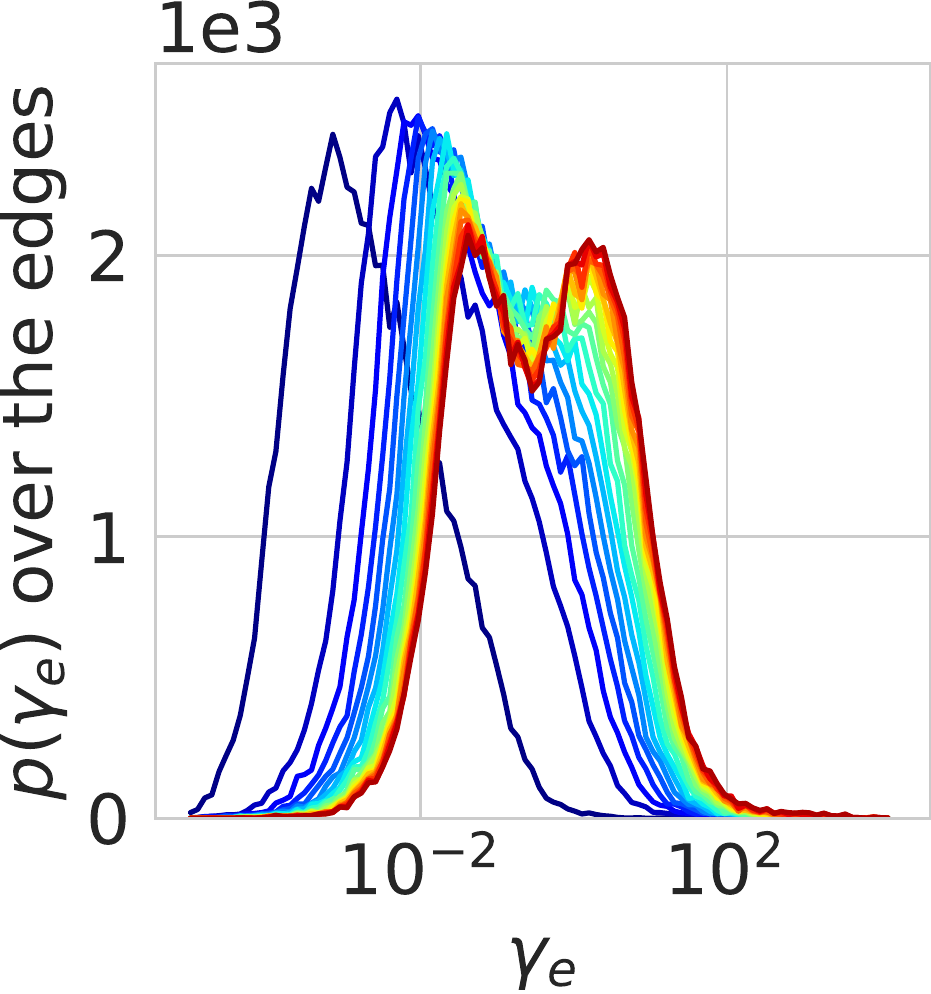}
      \end{subfigure}
      \begin{subfigure}{\linewidth/5 - 0.5em}
        \includegraphics[width=\linewidth]{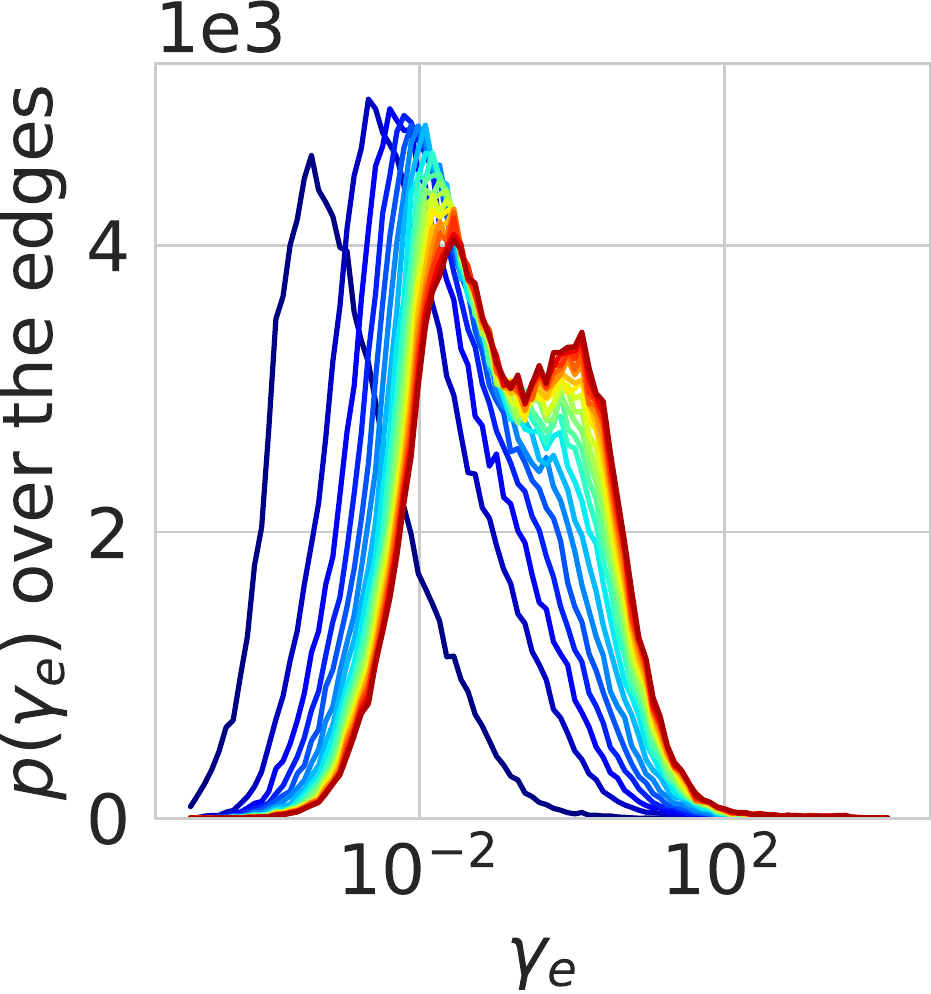}
      \end{subfigure}
      \begin{subfigure}{\linewidth/5 - 0.5em}
        \includegraphics[width=\linewidth]{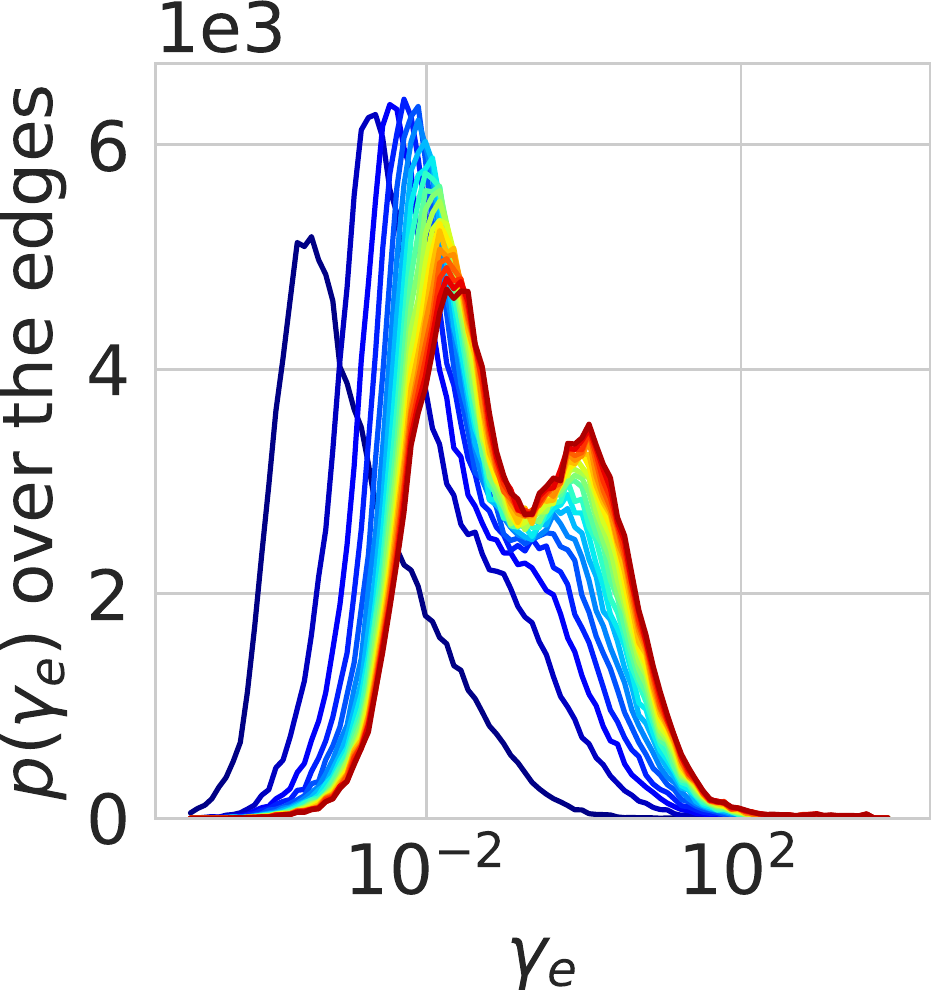}
      \end{subfigure}
    \end{subfigure}
    
    \par
    \begin{subfigure}{\linewidth}
    \centering
      \makebox[0pt]{\rotatebox[origin=c]{90}{
        (b) Time Spent
      }\hspace*{2em}}%
      \begin{subfigure}{\linewidth/5 - 0.5em}
        \includegraphics[width=\linewidth]{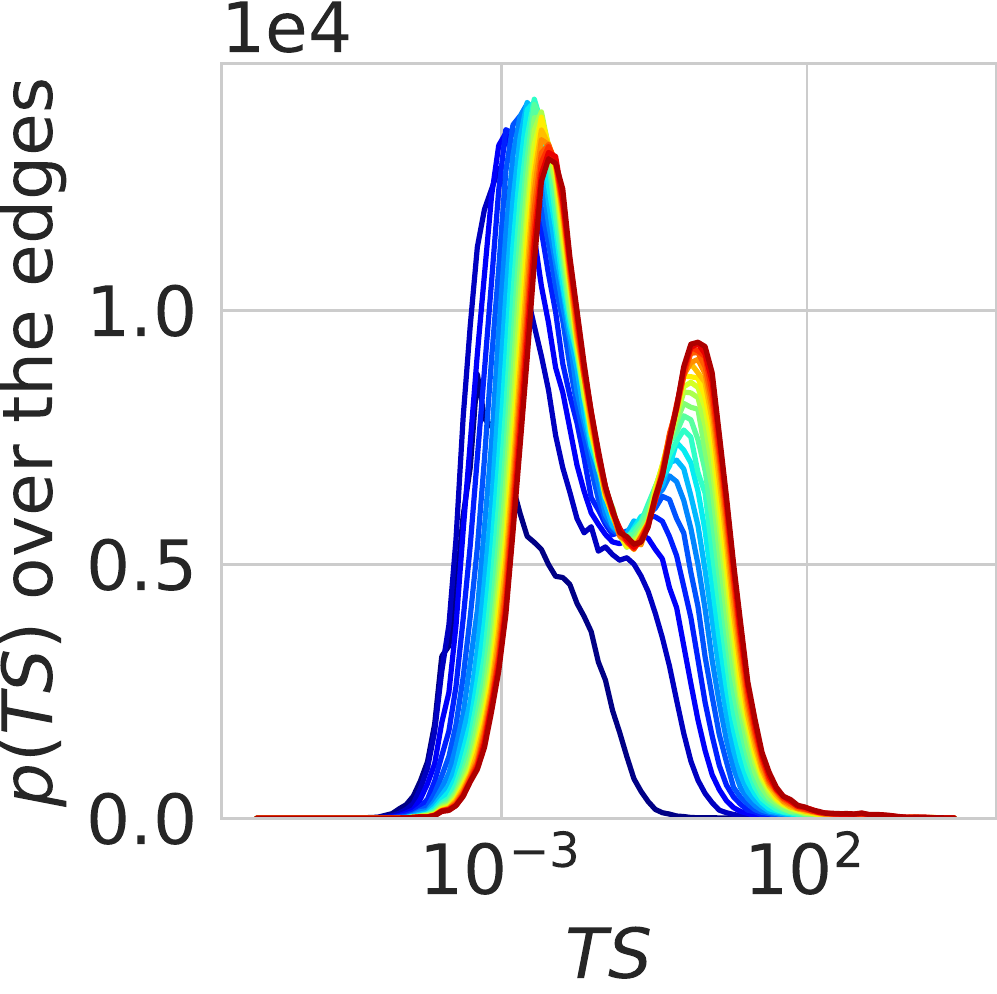}
      \end{subfigure}
      \begin{subfigure}{\linewidth/5 - 0.5em}
        \includegraphics[width=\linewidth]{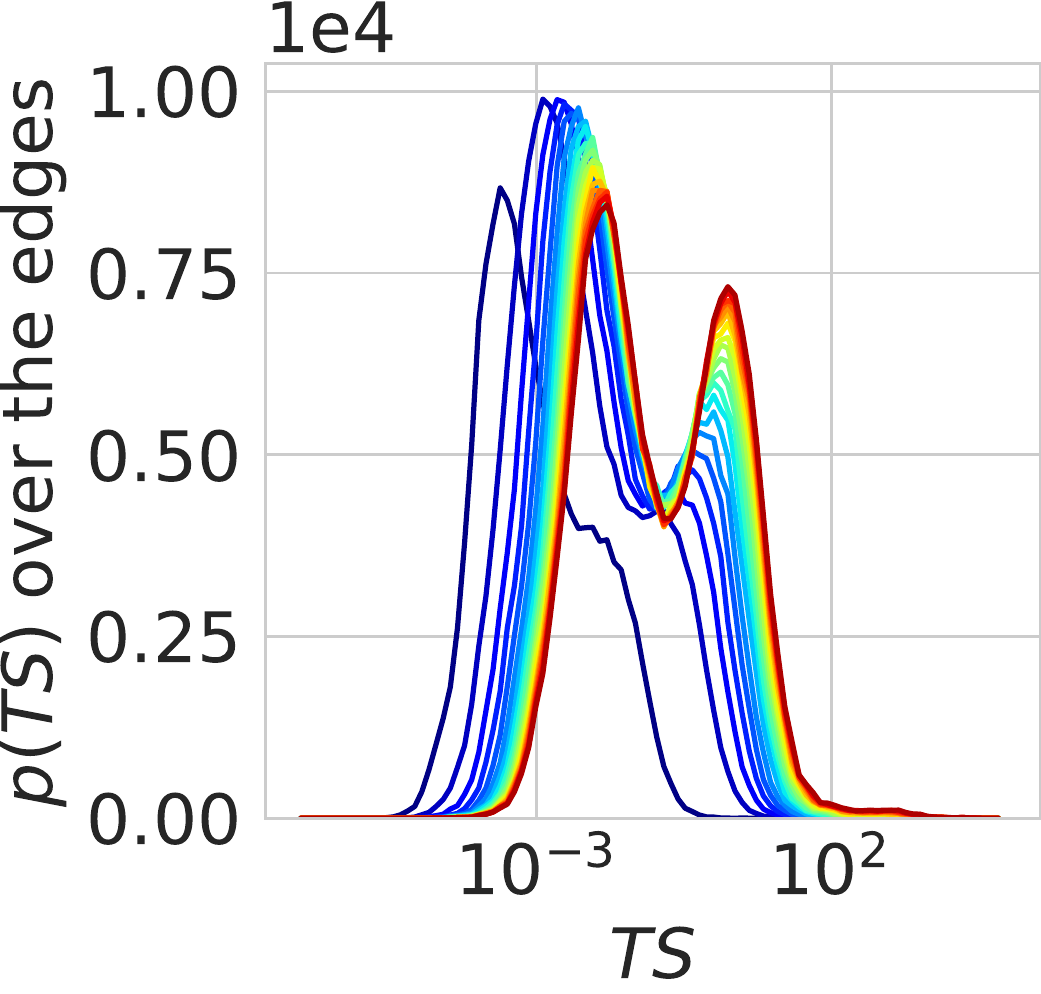}
      \end{subfigure}
      \begin{subfigure}{\linewidth/5 - 0.5em}
        \includegraphics[width=\linewidth]{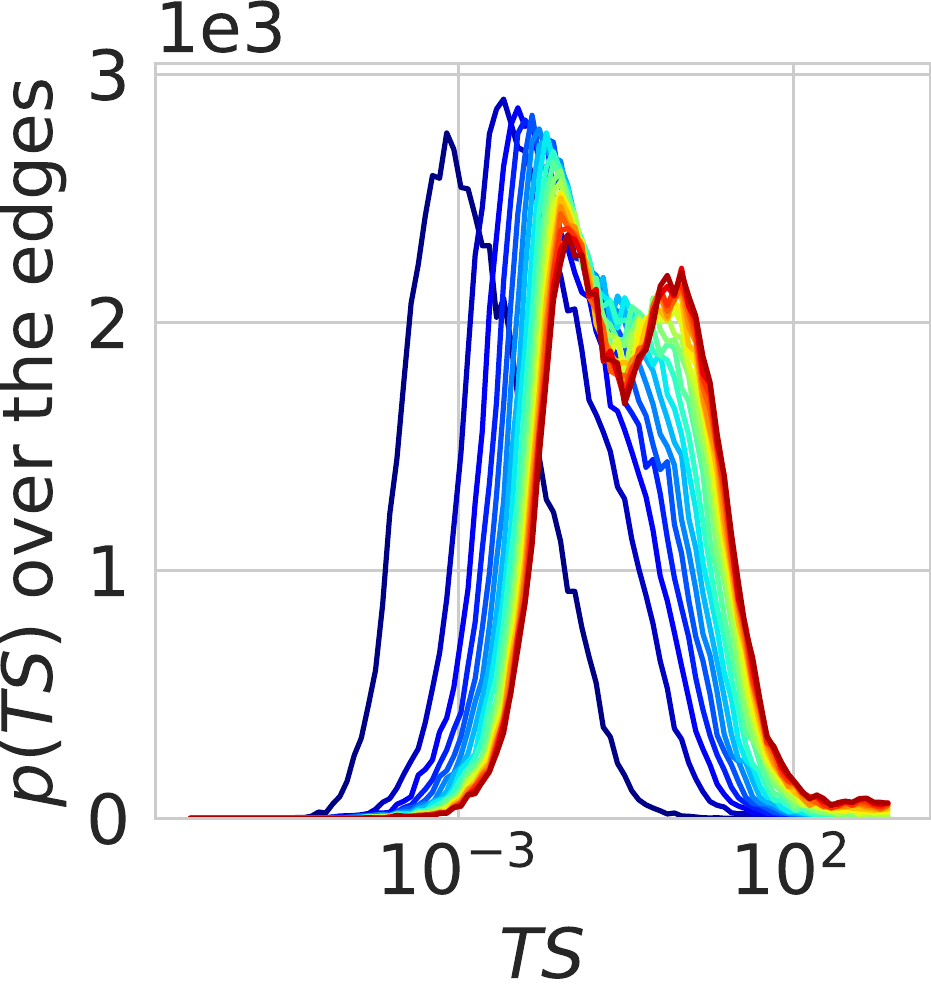}
      \end{subfigure}
      \begin{subfigure}{\linewidth/5 - 0.5em}
        \includegraphics[width=\linewidth]{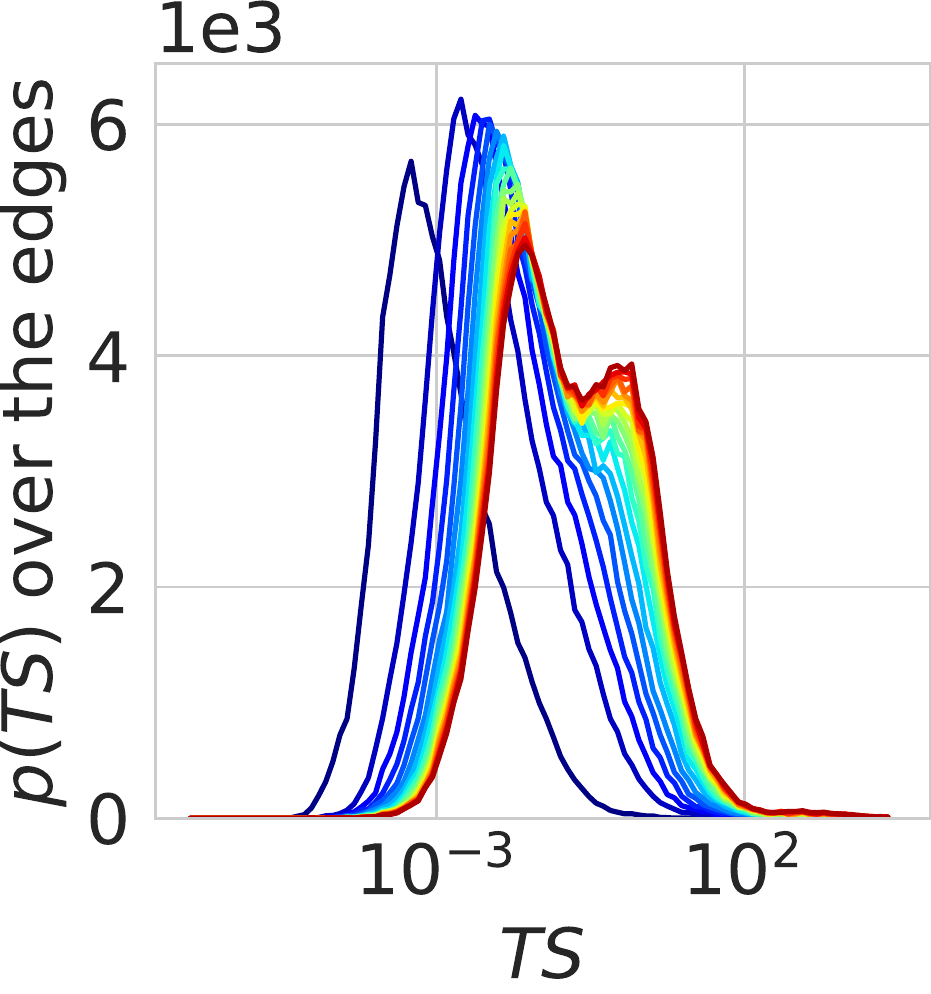}
      \end{subfigure}
      \begin{subfigure}{\linewidth/5 - 0.5em}
        \includegraphics[width=\linewidth]{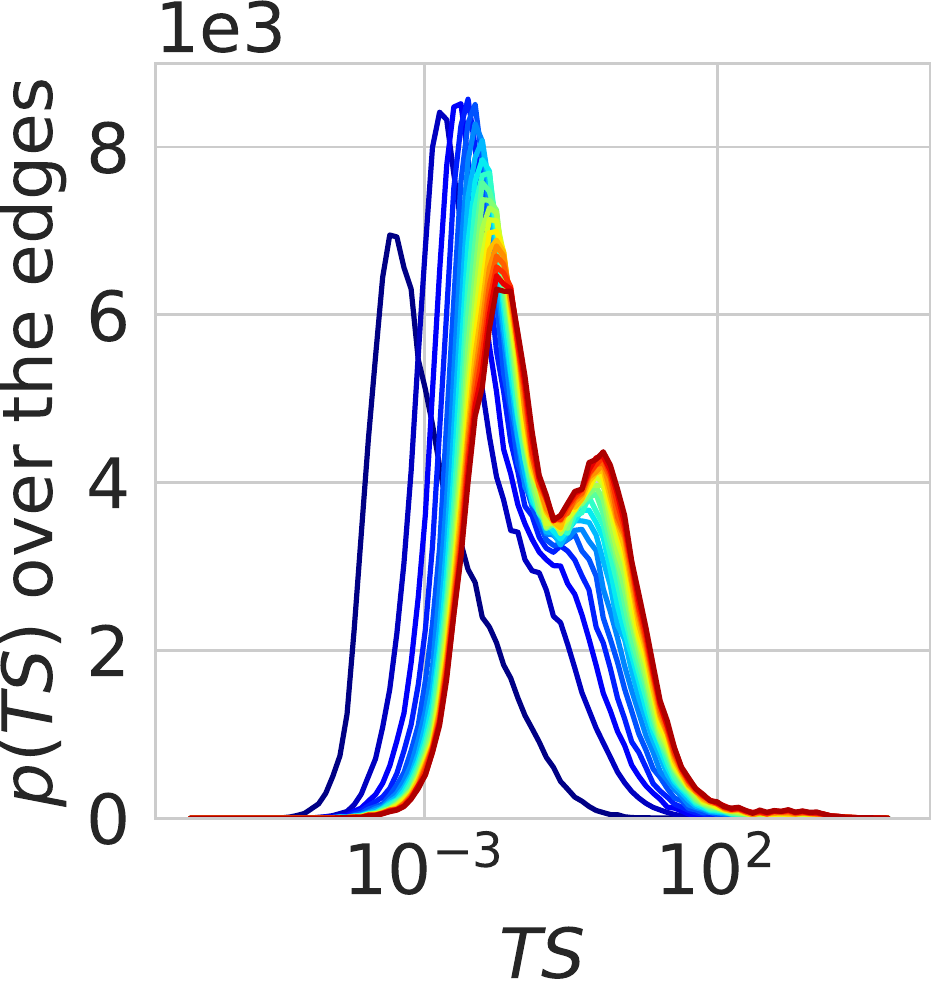}
      \end{subfigure}
    \end{subfigure}
   
    \par\vskip \abovecaptionskip
    \begin{subfigure}{\linewidth}
    \centering
      \setcounter{subfigure}{0}%
      \renewcommand\thesubfigure{\roman{subfigure}}
      \begin{subfigure}{\linewidth/5 - 0.5em}
        \caption{Boston}
      \end{subfigure}
      \begin{subfigure}{\linewidth/5 - 0.5em}
        \caption{London}
      \end{subfigure}
      \begin{subfigure}{\linewidth/5 - 0.5em}
        \caption{Rome}
      \end{subfigure}
      \begin{subfigure}{\linewidth/5 - 0.5em}
        \caption{Rio}
      \end{subfigure}
      \begin{subfigure}{\linewidth/5 - 0.5em}
        \caption{Nairobi}
      \end{subfigure}
    \end{subfigure}
\caption{Top: CBC [average occupancy per edge in 1h] for all cities at increasing levels of traffic (blue: light traffic, green: near transition, red: congested). Bottom: Time Spent in Traffic (TS) [vehicles$\cdot$hour]. Colors as for top row.}
\label{fig_matrix2}
\end{figure}

We study how the distributions of CBC (Fig.~\ref{fig_matrix2}(a)) and total time spent in traffic (TS) (Fig.~\ref{fig_matrix2}(b)) change for all cities when traffic grows from light to severely congested. 
As expected, both distributions shift to the right and a double peak structure starts to emerge already
for medium-low traffic levels. The indications of behavior change obtainable from the standard BC in the same
conditions are much weaker as seen in~\cite{kirkley_betweenness_2018} as the BC distribution is almost invariant
for totally different cities.

% figure for CBC and TS distributions
\onecolumngrid\
\twocolumngrid\

\section{Conclusion}
By extending the idea of BC, we proposed a novel approach based on a dynamical model to take
into account interactions among vehicles, to specifically characterize the peak-hour
network loading, typically the most demanding time in terms of infrastructural stress.
Our results show that the Cumulative BC is able to identify the bottlenecks of the network,
when subjected to a persistent load, and to study the back-propagation of traffic jams.
Interpreting the CBC as an expected road occupancy, we can also identify the edges responsible
for the largest contribution to the total time spent on average by all vehicles.
The results appear to be independent of OD order of addition for all the considered metrics,
but this may change with different dynamical models. In the future we plan to validate the
model against real traffic data (e.g., UBER Traffic Movement) and to improve it theoretically
(e.g., fine temporal evolution, different dynamical systems), while also extending it to wider
and more general networking contexts.

\section{Authors' Contributions}
M.~Cogoni and G.~Busonera contributed to the study conception and the
model design. All authors contributed to the writing and optimization
of the simulation code and to the analysis and interpretation of
results. All authors read and approved the final manuscript.

\section{Acknowledgments} This work has received funding from Sardinian Regional Authorities
under projects SVDC (art 9 L.R. 20/2015) and TDM (POR FESR 2014-2020 Action 1.2.2). Map data copyrighted OpenStreetMap contributors: www.openstreetmap.org

\clearpage

\bibliographystyle{unsrt}
\bibliography{traffic_paper_palermo2022}

\end{document}